\documentclass[a4paper,11pt]{article}
\pdfoutput=1 

\usepackage{jcappub} 

\usepackage[T1]{fontenc} 
\usepackage{color,soul}
\usepackage{placeins}
\usepackage{lscape}
\usepackage{longtable}

\usepackage{enumerate}

\title{\boldmath Informing dark matter direct detection limits with the \texttt{ARTEMIS} simulations}

\author[a]{Robert Poole-McKenzie,}
\author[a]{Andreea S. Font,}
\author[b]{Billy Boxer,}
\author[a]{Ian G. McCarthy,}
\author[b]{Sergey Burdin,}
\author[a]{Sam G. Stafford,}
\author[a]{Shaun T. Brown}

\affiliation[a]{Astrophysics Research Institute, Liverpool John Moores University, Liverpool, L3 5RF, UK}
\affiliation[b]{Oliver Lodge Laboratory, University of Liverpool, Liverpool, L7 7BD, UK}

\emailAdd{r.poolemckenzie@2013.ljmu.ac.uk}
\emailAdd{A.S.Font@ljmu.ac.uk}

\abstract{
Dark matter (DM) direct detection experiments aim to place constraints on the DM--nucleon scattering cross-section and the DM particle mass. These constraints depend sensitively on the assumed local DM density and velocity distribution function. While astrophysical observations can inform the former (in a model-dependent way), the latter is not directly accessible with observations. Here we use the high-resolution \texttt{ARTEMIS} cosmological hydrodynamical simulation suite of 42 Milky Way-mass halos to explore the spatial and kinematical distributions of the DM in the solar neighbourhood, and we examine how these quantities are influenced by substructures, baryons, the presence of dark discs, as well as general halo-to-halo scatter (cosmic variance). We also explore the accuracy of the standard Maxwellian approach for modelling the velocity distribution function. We find significant halo-to-halo scatter in the density and velocity functions which, if propagated through the standard halo model for predicting the DM detection limits, implies a significant scatter about the typically quoted limit.  We also show that, in general, the Maxwellian approximation works relatively well for simulations that include the important gravitational effects of baryons, but is less accurate for collisionless (DM-only) simulations.  Given the significant halo-to-halo scatter in quantities relevant for DM direct detection, we advocate propagating this source of uncertainty through in order to derive conservative DM detection limits.
}

\begin{document}
\maketitle 
\flushbottom

\section{Introduction}
\label{sec:intro}
Dark matter (DM) is the most important contributor to the mass budget in the Universe and plays a vital role in the formation of large-scale and galactic structures. Numerous particle candidates beyond the Standard Model (SM) have been proposed for DM. Among them, the so-called Weakly Interacting Massive Particle (WIMP) \cite{1985Steigman,2018Arcadi} has been studied extensively.  Definitive evidence for the existence of WIMPs, or of any other DM candidates, is actively sought via both direct and indirect detection experiments. In particular, direct detection experiments aim to detect WIMPs by measuring the nuclear recoil energy resulting from their elastic scattering off of atomic nuclei \cite{1985Goodman}. Indirect detection experiments, on the other hand, make use of either space- or ground-based telescopes to search for SM particles produced from the decay of WIMPs or from WIMP-WIMP annihilations that could occur within the Galaxy and/or in extragalactic sources (for a review see \citep{2018Arcadi}). In this study we focus on the predictions for the direct detection limits, using cosmological simulations. We focus on the potential signal from the solar neighbourhood in our Galaxy, which is deemed to be an important site for direct detection of DM.  

To date there is no conclusive evidence for direct detection of DM. While some positive signals have been reported from the DAMA/LIBRA \cite{2013Bernabei} and CDMS-II \cite{2013CDMS}, the evidence is not strong. At the same time, various null results have been reported from many other experiments, including the XENON1T \cite{2017AprileA}, which is one of the most sensitive experiments to date.

The predictions for the direct detection signals often assume the validity of the Standard Halo Model (SHM) \cite{1986SHM}. As the differential event rate in WIMP elastic scattering depends on both the local density and velocity distribution of WIMPs, the SHM involves certain assumptions about these distributions. From the assumption of a smooth, spherically symmetric DM halo, the velocity distribution of DM particles follows a Maxwell-Boltzmann function. The SHM also relies on observational measurements of several local Galactic parameters, such as the local DM density ($\rho_0$), circular velocity ($v_0$) and escape speed ($v_{\rm esc}$). Some of these measurements are still affected by systematic uncertainties (for example, $\rho_0$), and these measurements are inherently based on model assumptions for the Galaxy. 

Cosmological simulations can provide useful insights into some of the uncertainties in the SHM assumptions and how these propagate into direct detection limits \cite{2017Bozorgnia}. Both DM-only and hydrodynamical simulations have been used to study whether the DM velocities can deviate from a Maxwellian distribution. These deviations can occur, for example, when tidal streams cross the solar neighbourhood.  Deviations would result in the local velocity distribution being characterised by discrete peaks which, if ignored, could significantly bias the derived direct detection limits or measurements (e.g., \cite{2003Stiff}).  Interactions with other galaxies can move the Galaxy away from dynamical equilibrium and give rise to deviations from a Maxwellian distribution.

In this respect, some simulations suggest that the local DM velocity distribution can be significantly non-Maxwellian  \cite{2009Vogelsberger,2010Kuhlen,2010Ling,2011Lisanti,2013Mao,2014Pillepich,2016Sloane}. For example, \cite{2009Vogelsberger} analyse the Aquarius simulations \cite{2008Springel} and find secondary peaks in the velocity distribution of DM halos at $v \geq 250$ km s$^{-1}$, attributing these to the formation history of individual halos. Using a hydrodynamical simulation of a single Milky Way-sized halo, \cite{2010Ling} find that a Tsallis distribution \cite{1988Tsallis} best fits the velocity distribution. Using a sample of 96 halos simulated with hydrodynamics, \cite{2013Mao} find that the stacked velocity distribution has a wider peak and a steeper tail than a simple Maxwellian and suggest that the largest uncertainty in the velocity distribution arises from the radial position of the Milky Way with respect to the scale radius of the DM halo. In contrast, several other studies using cosmological hydrodynamical simulations find that the Maxwellian adopted in the SHM is a suitable approximation for the local velocity distribution. This has been shown, for example, by \cite{2016Kelso} for two Milky Way-mass halos from the \texttt{MaGICC} simulations \cite{2013Stinson}, and by \cite{2016Bozorgnia} for galaxies in the \texttt{EAGLE} \cite{2015SchayeB} and \texttt{APOSTLE} \cite{2016Sawala} simulations.

Note that the inclusion of baryons and associated physics in simulations may not only modify the local velocity distribution of the DM, but the baryons will also have a non-negligible impact on the spatial distribution of the DM, altering both its shape and its concentration (e.g., \cite{2008Read,duffy2010,deason2011,pontzen2012,schaller2015}). 

A dark disc can also potentially boost the DM signal in comparison with standard predictions using the SHM. Studies using hydrodynamical simulations have found that a dark disc has a negligible contribution  ($< 25$\%) to the local DM density and hence are not expected to contribute much to the DM direct detection signal \cite{2009Purcell,2010Ling, 2014Pillepich,2013Billard}. However, \cite{Read2009AGalaxies} found a wider range for the contribution of dark discs, of $(0.25 - 1.5) \times$ the non-rotating DM halo density near the Sun. A significant dark disc can increase the WIMP detection rates, e.g., by a factor of $\approx 3$ at recoil energies of $5-20$~keV \cite{2009Bruch} and thus improve the constraints on the interaction cross-section. 

In this study, we aim to re-evaluate the spatial and kinematical distribution of local DM and to examine the prevalence of dark discs and of local substructure using a new set of high-resolution hydrodynamical simulations, as well as to estimate how these structures can potentially impact the DM direct detection limits. The new suite of high resolution, zoomed-in simulations, called \texttt{ARTEMIS}, follows the growth of $42$ Milky Way-mass galaxies in a $\Lambda$CDM cosmological model.  Each halo in the suite has two realisations: a collisionless version (hereafter DMO) and a fully hydrodynamical version (hereafter `hydro').  \texttt{ARTEMIS} is the largest sample to date for this type of prediction at such high resolution (with baryon and dark matter simulation particle masses of $2.2\times10^4$ M$_\odot\/h^{-1}$ and $1.2\times10^5$ M$_\odot\/h^{-1}$, respectively).  As we discuss below, and as previously shown in \cite{2020Font}, the \texttt{ARTEMIS} simulations reproduce the observed stellar masses and disc sizes of Milky Way-mass galaxies remarkably well and should therefore realistically capture the gravitational impact of the baryons on the DM (and vice-versa).

This large suite of simulations allows us to investigate not only the impact of baryons in a realistic way, but also allows us to assess the impact of halo-to-halo scatter on the predictions.  Furthermore, by using the local density and local velocity distributions of DM from these simulations and their scatter, we can inform the detection limits in direct detection experiments such as LUX-ZEPLIN (LZ) \cite{LZ_2018} and XENON1T.  In practice, this means that the existing detection limits actually turn into fuzzy bounds when one propagates the halo-to-halo scatter through, implying that the limits themselves have non-negligible uncertainties. 

The paper is organised as follows. In Section~\ref{sec:methods}, we introduce the set of high resolution, cosmological hydrodynamical simulations of Milky Way-mass galaxies that will be used in our study. We also briefly discuss the formalism of the SHM and of predicting the direct detection signal. In Section~\ref{sec:local} we determine the range of local DM densities in the simulated halos, the distribution function of DM velocities in their solar neighbourhoods and estimate the impact that substructure may have locally. In Section~\ref{sec:non-local} we investigate other (non-local) changes in the structure of DM, including changes in the DM halo shapes, and we search for evidence of dark discs in these halos. In Section~\ref{sec:cross-section}, we show how the variations in the local DM properties may affect the predicted exclusion limits of DM direct detection experiments, namely those for LZ and XENON1T. We also adopt an empirical model for the local velocity distribution based on our simulations and show how it compares to the SHM predictions. Finally, in Section~\ref{sec:Conclusion}, we summarise our findings and conclude.

\section{Methods}
\label{sec:methods}
\subsection{The \texttt{ARTEMIS} simulations}
\label{sec:sims}

We use the new \texttt{ARTEMIS} suite of high-resolution cosmological hydrodynamical simulations of Milky Way-mass halos \cite{2020Font}.  Full details of the simulations are provided in \cite{2020Font}, but we provide an overview of the simulations here.

The \texttt{ARTEMIS} simulations employ the `zoom in' technique (e.g.~\cite{bertschinger2001}) to simulate Milky Way-analog halos at high resolution and with hydrodynamics, within a larger box that is simulated at comparatively lower resolution and with collisionless dynamics only.  The initial conditions were generated using the \texttt{MUSIC} code\footnote{\url{https://www-n.oca.eu/ohahn/MUSIC/}} \cite{hahn2011}. Halos were selected from a base periodic box is $25$~Mpc $h^{-1}$ on a side with $256^3$ particles. The initial conditions were generated at a redshift of $z=127$ using a transfer function computed using the \texttt{CAMB}\footnote{\url{https://camb.info/}} Boltzmann code \cite{lewis2000} for a flat $\Lambda$CDM WMAP9 \cite{hinshaw2013} cosmology ($\Omega_\textrm{m}=0.2793$, $\Omega_\textrm{b}=0.0463$, $h=0.70$, $\sigma_8=0.8211$, $n_s=0.972$), which we adopt here.  The initial conditions include second order Lagrangian perturbation theory (2LPT) corrections.

The base periodic volume was run down to $z=0$ using the Gadget-3 code (last described in \cite{springel2005}) with collisionless dynamics.  Milky Way analogs were selected based on total halo mass; specifically, \cite{2020Font} selected a volume-limited sample of halos (i.e., all halos) whose total mass fell in the range $8\times10^{11} < {\rm M}_{200,{\rm crit}}/{\rm M}_\odot < 2\times10^{12}$, where ${\rm M}_{200,{\rm crit}}$ is the mass enclosed inside a sphere with radius $\rm{R}_{200,{\rm crit}}$, when the mean density is 200 times the critical density at $z=0$. 
This approximately spans the range of values inferred for the Milky Way from a variety of different observations, i.e, ${\rm M}_{200} \approx 0.55 - 2.62 \times 10^{12} \, {\rm M}_{\odot}$ \cite{2016Bland-Hawthorn}. There are $63$ such halos in this mass range in the periodic volume.  In the present study, we use a subset of 42 high-resolution collisionless simulations (DMO), together with their full hydrodynamical counterparts, presented in \cite{2020Font}. We note that the subset of 42 halos was not explicitly selected based on any physical criterion.  They are the subset of halos that managed to run to $z=0$ in the allotted HPC time allocation.  Because of the nature of the halo mass function and that higher-mass halos tend to be more computationally expensive at fixed resolution, the high-mass end of the initial range is not particularly well sampled in the completed subset of 42. The resulting median (mean) halo mass of the subset is 
${\rm M}_{200,{\rm crit}} \approx 1.01 \ (1.11) \times 10^{12}$ M$_\odot$.  A consequence of this is that the peak circular velocities tend to be on the lower side of that observed for the Milky Way, though there is some overlap.

The zoomed ICs were generated by first selecting all particles within $2 {\rm R}_{200,{\rm crit}}$ of the selected halos and tracing them back to the initial conditions of the periodic box, at $z=127$, to define the region which would be re-simulated at higher resolution and (for the hydro simulations) with baryons.
The outer radius for particle selection was chosen to ensure that we simulate, at high resolution, a region that at least encloses the splashback radius, which marks the physical boundary of the halo out to which particles pass on first apocenter \cite{diemer2014}.  

The base periodic run has a \texttt{MUSIC} refinement level of 8, whereas the zoom region has a maximum refinement level of $11$.  With this level of refinement, the DM particle mass is $1.17\times10^5$ M$_\odot\/h^{-1}$ and the initial baryon particle mass is $2.23\times10^4$ M$_\odot\/h^{-1}$.  Following the convergence criteria discussed in \cite{power2003}, a force resolution (Plummer-equivalent softening) of $125$~pc/$h^{-1}$ (which is in physical coordinates below $z=3$ and comoving coordinates at earlier times) was adopted.

Note the resolution of \texttt{ARTEMIS} is similar to that of the highest resolution simulations from other groups for this mass scale.  For example, \texttt{ARTEMIS} lies between resolution levels 3 and 4 (with 3 the highest) of the Auriga simulations \cite{grand2017} and levels 1 and 2 (1 being the highest) of the \texttt{APOSTLE} simulations \cite{2016Sawala}, which also uses the \texttt{EAGLE} code. It is also comparable in resolution to the \texttt{FIRE-2} simulations of Milky Way-analog halos \cite{garrison-kimmel2018}.  However, in general the \texttt{ARTEMIS} sample is larger, in terms of the number of Milky Way analogs simulated at this very high resolution, and provides us with the opportunity to explore the uncertainties due to cosmic variance (i.e., halo-to-halo scatter) in the predictions.

To carry out the hydrodynamical zoomed simulations, \texttt{ARTEMIS} uses the Gadget-3 code with an updated hydro solver and galaxy formation modelling (subgrid physics) developed for the \texttt{EAGLE} project \cite{schaye2015}.  The \texttt{EAGLE} model includes subgrid prescriptions for important processes that cannot be resolved directly in the simulations, including metal-dependent radiative cooling, star formation, stellar evolution and chemodynamics, black hole formation and growth through mergers and gas accretion, along with stellar feedback and feedback from active galactic nuclei (AGN) (see \cite{schaye2015} and references therein).

An important consideration for galaxy formation modelling is the calibration of the feedback efficiencies.  At the scale of Milky Way-mass galaxies, stellar feedback is expected to dominate over that of AGN.  \cite{2020Font} adjusted the parameter values of the stellar feedback model in the \texttt{EAGLE} code to reproduce the amplitude of the observed galaxy stellar mass--halo mass relation at the Milky Way halo-mass scale.  This means that, at a given halo mass, the simulations have realistic stellar masses compared to the global galaxy population, by construction. This is crucial for the current study, as the simulations should, as a result, realistically include the gravitational impact of baryons on the underlying DM distribution (and vice-versa).  While the simulations were not calibrated on other aspects of the observed galaxy population, they nevertheless reproduce a number of key observables, including the disc size--stellar mass and star formation rate--stellar mass relations.  

As noted above, the halo masses (and circular velocities) of the subset of 42 haloes tend to be on lower side of the allowed halo mass range for the Milky Way. Consequently, the stellar masses are typically also somewhat low compared to that quoted for the Milky Way (see \cite{2020Font} for discussion).  Nevertheless, there is still some overlap between the \texttt{ARTEMIS} subset of 42 and the observed stellar mass of the Milky Way.  For example, \cite{2016Bland-Hawthorn} estimate a total stellar mass for the Milky Way of $5\pm1 \times 10^{10}$ M$_\odot$, while \texttt{ARTEMIS} ranges from $(1.75 - 5.45) \times 10^{10}$ M$_\odot$, with a mean (median) stellar mass of $2.87 \ (2.96) \times 10^{10}$ M$_\odot$.

It is important to note that there is a non-negligible ($\approx0.2$ dex; e.g., \cite{2013Behroozi}) intrinsic scatter in the empirical stellar mass--halo mass relation of galaxies.  This implies that, even if the simulations were to recover the stellar mass of the Milky Way perfectly (although, as noted above there is still considerable uncertainty in the observed mass), this would not automatically imply that the simulations would have the correct halo mass and/or circular velocity for the Milky Way.  Indeed, we find that there is considerable scatter in the circular velocities at fixed stellar mass.

Given these uncertainties, our approach is to concentrate on the scatter in the implied DM detection limits for a sample of approximately fixed halo mass, and to examine the relative effects of hydro simulations to DM only simulations.  We acknowledge that an alternative way to proceed would be to select simulated galaxies of approximately fixed stellar mass (ideally one consistent with estimates of the Milky Way) and explore the scatter that results from the scatter in halo mass and circular velocities.  Given the nature of the halo mass-selected \texttt{ARTEMIS} sample, though, we must leave this for future work.

Returning to the discussion of the sample, the DMO halos have been matched with halos from the hydro simulations by using the unique particle IDs of the DM particles.  By uniquely matching halos from the collisionless DMO simulations to those in the hydro simulations, we can unambiguously determine the impact of baryons on the DM spatial and velocity distributions. The global properties for the halos in both the DMO and hydro simulations can be found in Table~\ref{tab:halo_properties} in Appendix \ref{sec:appendix}.

\begin{figure}[!ht]
  \includegraphics[width=\textwidth]{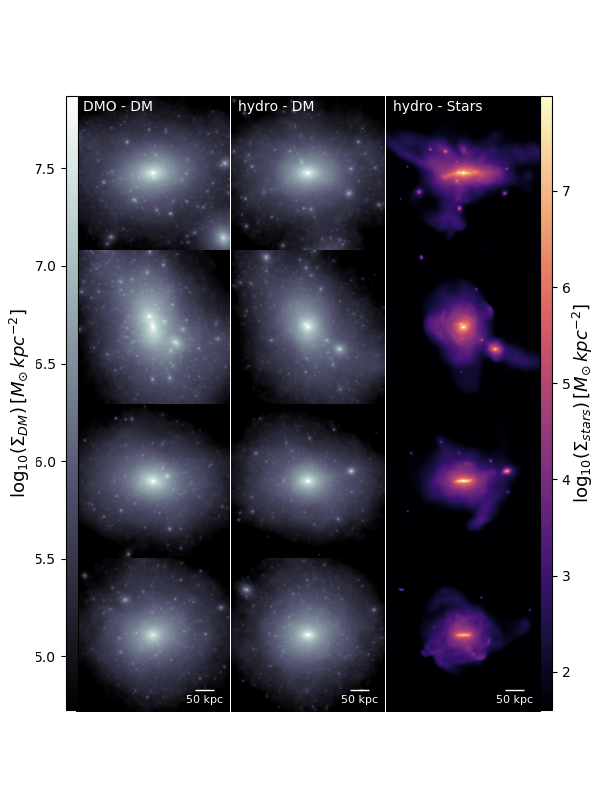}
  \caption{\textit{Left column}: The projected density field of DM in the DMO simulations for Milky Way-mass halos: G25, G28, G34, G38 (top to bottom). Numerous subhalos can be observed in these distributions. \textit{Middle column}: The projected density field of DM for the same halos in the hydro simulations. \textit{Right column}: The edge-on projected density field of the star particles in the same four galaxies in the hydro simulations. A variety of stellar streams can be seen in addition to gravitationally bound satellite galaxies.}
  \label{fig:halos}
\end{figure}

Four examples of halos from the sample used in this study are shown in Fig.~\ref{fig:halos}, displaying the present-day, projected DM density fields in the DMO halos and in the matched halos in the hydro simulations, alongside stellar density maps of the same systems as viewed edge-on to the disc components.  

\subsubsection{Specifying simulated `solar neighbourhoods'}

When analysing the simulations, we adopt a `Galactic' coordinate system.  Velocities are expressed with respect to the centre of mass velocity of the halo, and the halo centre is chosen to be the centre of potential (i.e., the location of the most bound particle). For the hydro simulations, the stellar discs are identified with the kinematical method outlined in \cite{2020Font}. For each Milky Way-mass galaxy, the $z$ axis is chosen to lie along the direction of the total angular momentum of stellar particles in the inner $20$~kpc region of each system. The $x$ and $y$ axes are therefore in the plane of the stellar disc. For consistency, we use the same system of reference for the DMO halos (which have no stellar discs) as determined in their matched hydro counterparts. 

With a reference frame established, we select `solar neighbourhood' regions for each simulation. Given that the simulated sample covers a range in virial masses and radii (with radii ranging from ${\rm R}_{200,{\rm crit}} \approx 180 - 250$~kpc, see Table~\ref{tab:halo_properties}), and because the stellar discs differ in terms of their scalelengths, we take `solar radius' in each simulation to be a fixed fraction of ${\rm R}_{200,{\rm crit}}$, specifically ${\rm R}_0 = 0.04 \, {\rm R}_{200,{\rm crit}}$ for each system.
This compares well with the scaling of ${\rm R}_0/{\rm R}_{200}$ for the Milky Way, which has an estimated virial radius, ${\rm R}_{200,{\rm crit}}$, between $168.7-283.4$~kpc and an estimated solar Galactocentric radius, ${\rm R}_{\odot}$, between $7.10-8.92$~kpc (for both, see \cite{2016Bland-Hawthorn} and references therein), the measurements for the latter showing considerable scatter around the `standard' value of $8.5$~kpc. Furthermore, we consider the solar neighbourhood to be a cylindrical shell with a radial distance of ${\rm R}_0$, a fixed width of $1$~kpc and a fixed height of $1.5$~kpc. (We have investigated that changing the width or height of the `solar neighbourhood' region or adopting a fixed distance, e.g., ${\rm R}_0 = 8.5$~kpc does not significantly change our results).

With these parameters, the cylindrical shells contain a substantial number of DM particles, ranging from $\approx8,800-22,500$  for the hydro simulations and between $\approx5,400-18,000$ in the DMO simulations.  (The larger number of DM particles in the hydro simulation is due to adiabatic contraction of the DM due to the presence of the baryons, as we discuss later.)

\FloatBarrier
\FloatBarrier

\subsection{Standard Halo Model}
\label{sec:SHM}

The SHM \cite{1986SHM} is routinely used in the modelling of data from direct detection experiments, or in making predictions for such detections. The model assumes a simple spherically-symmetric DM profile (usually either an isothermal or a Navarro-Frenk-White profile \citep{NFW1997}) corresponding to a halo with a total mass equal to that of the Milky Way.  This model leads to a Maxwell-Boltzmann distribution of velocities for DM particles in the Galactic frame, $f(\vec{v}),$ which, in order to account for the finite size of the halo, is truncated at the escape velocity:

\begin{equation}
f(\vec{v}) = \frac{1}{(2\pi \sigma^2_v)^{3/2} N_{\rm esc}} \,  \exp{\Big(-\frac{\vert \vec{v} \vert^2}{2 \sigma^2_v}\Big) \, \, \Theta(v_{\rm esc} - \vert \vec{v} \vert)},
\end{equation}
    
\noindent where $\sigma_v$ is the velocity dispersion of the DM, which is related to the most probable DM velocity (taken to be the local circular velocity), $v_0$, via $\sigma_v = v_0/\sqrt{2}$, and $\Theta$ is the Heaviside function that truncates the distribution.  As the integral over the velocity dispersion needs to be unity for the calculation of the scattering rates (see below), the Maxwellian must be renormalised to account for the truncation via the parameter $N_{\rm esc}$, defined as:

\begin{equation}\label{nesc}
N_{\rm esc} = {\rm erf}(z) - 2z\, \exp(-z^2)/\pi^{1/2} \ ,
\end{equation}
\noindent where \({\rm erf}()\) is the error function and \(z = v_{\rm esc}/v_0\).

In practice, we truncate the Maxwellian function based on certain conditions of the WIMP velocities \cite{Lewin1996ReviewRecoil}:

\newcommand{\threepartdef}[6]
{
	\left\{
		\begin{array}{ll}
			#1 & \mbox{if } #2 \\
			#3 & \mbox{if } #4 \\
			#5 & \mbox{if } #6
		\end{array}
	\right.
}

\newcommand{\firstPart}{\frac{1}{v_0y}}

\newcommand{\secondPart}{\frac{1}{2N_{\rm esc}v_0y}\left[{\rm erf}(x+y)-{\rm erf}(x-y)-\frac{4}{\sqrt{\pi}}ye^{-z^2}\right]}

\newcommand{\thirdPart}{\frac{1}{2N_{\rm esc}v_0y}\left[{\rm erf}(z)-{\rm erf}(x-y)-\frac{2}{\sqrt{\pi}}(y+z-x)e^{-z^2}\right]}

\begin{equation}\label{eq:fv}
\small
\hspace*{-2cm}
\int_{v_{\rm{min}}}^{v_{\rm{max}}} \frac{f(\vec{v})}{v} =
\threepartdef { \firstPart } {z<y,\ x<|y-z| \\} {\secondPart} {z>y, \ x<|y-z| \\} {\thirdPart} {|y-z| < x < y+z }
\end{equation} 

\noindent where \(x = v_{\rm{min}}/v_0\), \(y=|V_{E}|/v_0\) (where \(|V_{E}|\) is the velocity of the detector frame within the halo frame), and $z$ is as defined above.  Note that the first condition of the integral is not achievable in practice, as for this to be the case the Earth's velocity would have to be greater than the escape velocity.  It is only included here for completeness. 

To fully specify the truncated Maxwellian above, only two parameters are required: $v_0$ and $v_{\rm esc}$.  For $v_0$ it is standard practice to adopt the rotational speed of the Sun around the centre of the Milky Way (typically assumed to be $220$~km $\rm{s^{-1}}$), assuming it reflects the local circular velocity of the Galaxy. However, the observational values for the latter vary, and a more recently revised value is $238 \pm 15$~km s$^{-1}$ \cite{2016Bland-Hawthorn}.

Estimates of the escape speed come from measurements of high-velocity stars in the solar neighborhood. Several experiments, including LZ and XENON1T \cite{LZ_2018,2018Aprile}, use a value of $v_{\rm esc}=544$ km s$^{-1}$. As discussed by \cite{2019Evans}, this value is based on an earlier measurement from the RAVE survey that used only 12 high velocity stars. Using a slightly larger sample of stars from the same survey, \cite{2014Piffl} obtain a value of $533_{-41}^{+54}$ km s$^{-1}$. More recently, this value has been revised  using data from the \textit{Gaia} survey. While, initially, this has led to a higher value, of $580 \pm 63$ km s$^{-1}$ \citep{2018Monari}, a subsequent analysis has obtained $528_{-25}^{+24}$ km s$^{-1}$ \cite{Deason2019}. 

In addition to a velocity distribution, the local DM density is required to compute the expected scattering rates.  Estimates of the local WIMP density come from a range of sources, including the use of local dynamical estimates applied to stars in the solar neighbourhood, which must make assumptions about the geometry and state of equilibrium of the underlying DM component.  Typically, the local DM density is taken to be $\rho_0 = 0.3$ GeV cm$^{-3}$ \cite{2017Green, Read2014TheDensity}. However, recent observational results from stellar kinematics, stellar density profiles, maser observations and gas velocities suggest that a larger range of values of $\rho_0$ $\sim 0.45 - 0.70$ GeV cm$^{-3}$ may be more appropriate \citep{2012Smith, 2014Bienaym, 2018Sivertsson, 2018Hagen, 2014MNRAS.445.3133P}. Predictions for $\rho_0$ from cosmological simulations, as we produce here, will depend on the assumed mass of the Milky Way, for which measurements still have relatively large uncertainties \cite{2016Bland-Hawthorn}. This uncertainty is, to an extent, incorporated in our analysis, as we select halos whose masses span the range of quoted values for the Milky Way. 

Direct DM detection aims to measure the nuclear recoil of a SM particle as it interacts with a WIMP. The differential scattering rate of a WIMP-nuclei interaction depends on the local DM density and the velocity distribution and can be written as:

\begin{equation}
\frac{dR}{dE}(E,t)=\frac{\rho_0}{m_{\rm DM}\, m_N}\int_{v_{\rm{min}}}^{v_{\rm esc}}vf_E(\vec{v})\frac{d\sigma}{dE}(v,E)d^3\vec{v},
\label{eq:DiffSR}
\end{equation}
 
\noindent where $\rho_0$ is the local DM density, m$_{\rm DM}$ and m$_N$ are the DM and nuclei particle masses (respectively), $v_{\rm min}$ is the minimum velocity the particle requires to produce a detection at the recoil energy $E$, $\vec{v}$ is the velocity vector of the DM particle relative to the Earth, $f_E(\vec{v})$ is the corresponding velocity distribution function, and $d\sigma/dE$ is the energy differential DM-nucleus scattering cross-section.  The minimum velocity depends to the threshold recoil energy in the form:

\begin{equation}
v_{\rm min} = \sqrt{\frac{E m_N}{2 \mu_{{\rm DM},N}^2}},
\label{eq:vmin}
\end{equation}
\noindent where $\mu_{{\rm DM},N} \equiv (m_{\rm DM} m_N) / (m_{\rm DM}+m_N)$ is the DM-nucleus reduced mass.

For our analysis, we use a simplified version of equation \ref{eq:DiffSR}, specifically: 

\begin{equation}
\frac{dR}{dE} = \frac{\rho_0\sigma_0}{2m_{\rm DM}\mu_{{\rm DM},N}^2}F(q)^2\int_{v_{\rm{min}}}^{v_{\rm esc}} \frac{f(\vec{v})}{v} d^3 v,
\label{eq:BBRate}
\end{equation}

\noindent where $\sigma_0$ is the zero momentum interaction cross-section, $F(q)^2$ is the nuclear form factor (which is a measure of the scattering amplitude of an incoming particle of a signal atom) and $q \equiv \sqrt{2 m_N E}$.

The spin-independent form factor is taken to be \cite{1991Engel}:

\begin{equation}
F(q)^2 = \biggl(\frac{3 j_i(qr_n)}{qr_n}\biggr)e^{-q^2 s^2},
\end{equation}

\noindent where $j_i$ is the Bessel function (for which we use only the first order version; i.e., $j_1$), $r_n$ is the reduced nucleon radius and $s$ is the nucleon skin depth ($\sim$ 1 fm).

As can be inferred from eqns.~\ref{eq:DiffSR} and \ref{eq:BBRate}, the DM-nucleon scattering rate is highly dependent on the assumed velocity distribution function and density of the DM.  Below, we examine these quantities in the \texttt{ARTEMIS} simulations and how they are influenced by substructures, baryons, the presence of dark discs, as well as general halo-to-halo scatter (cosmic variance).

\section{Local dark matter distributions in \texttt{ARTEMIS}}
\label{sec:local}

\subsection{Local density and velocity distributions}
\label{sec:localdens_vel}

\begin{figure}[!htb]
    \centering
    \includegraphics[width=\textwidth]{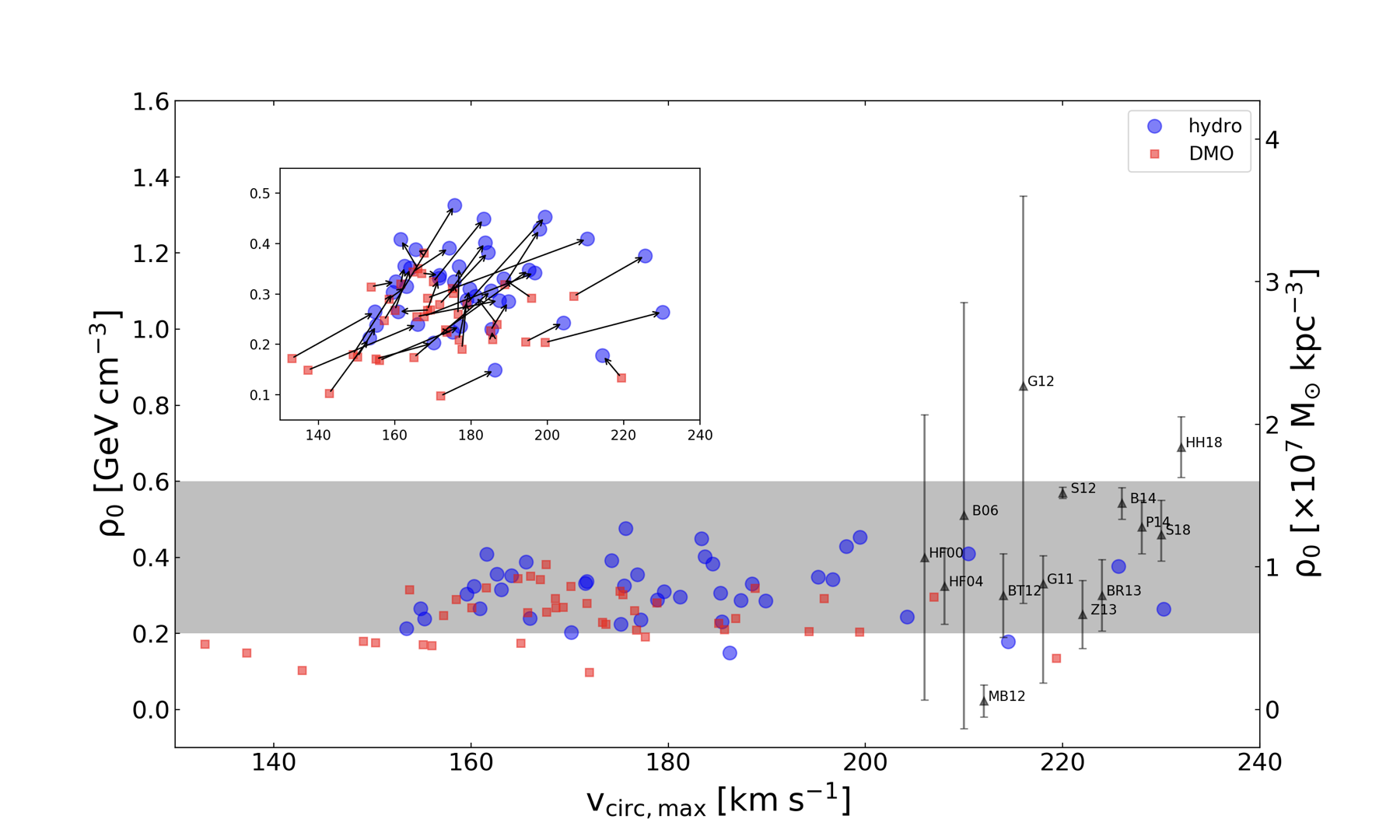}
    \caption{The local DM density, $\rho_0$ versus maximum circular velocity, $v_{\rm{circ,max}}$, for the halos in the hydro (blue-filled circles) and DMO (red-filled squares) simulations. The black triangles represent local measurements of $\rho_0$ and associated errors (see text for details; points have been shifted in $v_{\rm{circ,max}}$ around the $220$ km s$^{-1}$ value, for readability). The grey band represents the range of additional estimates of $\rho_0$ from \cite{2017Green}. \textit{Inset panel:} A zoom-in on the $\rho_{0}$ versus $v_{\rm{circ,max}}$ plot for the simulated systems. The arrows show the shift in the ($\rho_0$, $v_{\rm{circ,max}}$) values from the DMO halos to their matched hydro counterparts.}
    \label{fig:LocDen}
\end{figure}

Fig.~\ref{fig:LocDen} shows the distribution of the local DM densities (averaged over the cylindrical shell), $\rho_0$, versus the maximum circular velocities, $v_{\rm circ, max}$, for all  $42$ Milky Way-mass systems. Blue-filled circles represent the values in the hydro simulations and red-filled squares, those in the DMO simulations. We also plot observational measurements of the local DM density in the Milky Way, shown as black triangles and their respective error bars \cite{2004Holmberg, 2006Bienayme, 2012Moni-Bidin, Bovy2012OnDensity, 2011Garbari, 2012Garbari, 2012Smith, 2013Zhang, 2013Bovy, 2014Bienaym, 2014MNRAS.445.3133P, 2018Hagen, 2018Sivertsson}. The grey band shows the range of other $\rho_0$ measurements (see \cite{2017Green} and references therein). These indicate that there is a still significant uncertainty in the value of $\rho_0$, due to the variety of observational methods and implicit assumptions in the modelling \cite{Pato2010SystematicDensity, Read2014TheDensity}. The errors are known to be dominated by systematic effects, as indicated by the scatter in these values being larger than the errors of individual measurements. 

Overall, the $\rho_0$ values obtained from our simulations (both DMO and hydro) agree reasonably well with the many of the observational constraints, which is reassuring given the very different method we have employed to estimate the density.  Specifically, the median $\rho_0$ in the hydro simulations is 0.32 GeV cm$^{-3}$, with a full range of $0.15 - 0.48$ GeV cm$^{-3}$, while in the DMO simulations the median is 0.26 GeV cm$^{-3}$ with a range of $0.10 - 0.38$ GeV cm$^{-3}$. The median values agree well with the most quoted value of $0.3$ GeV cm$^{-3}$. Our simulations clearly disfavour the higher values found in some observations (i.e., $\rho_0 >0.6$ GeV cm$^{-3}$), even after when baryonic effects are taken into account.

Interestingly, we find that there are marked differences between the local DM densities in the hydro halos and their respective DMO counterparts. For clarity, the inset panel in Fig.~\ref{fig:LocDen} shows a zoom into the $\rho_{0}$ and $v_{\rm{circ,max}}$ values, with the arrows showing the direction of the changes between the DMO and hydro simulations. Generally, both $\rho_0$ and $v_{\rm{circ,max}}$ increase for the same halo in the presence of baryons, which is due to the adiabatic contraction of the DM halo in response to the baryons \cite{1986Blumenthal, 2004Gnedin}. This implies that the DMO simulations do not capture all the important physical processes necessary for predicting $\rho_0$ or $v_{\rm{circ,max}}$, and that hydrodynamical simulations are better suited for this task (so long as the stellar mass distributions are realistic, as in \texttt{ARTEMIS}).  We investigate the impact that these effects have on DM detection limits in Section~\ref{lz_xenon1T_SHM}.

\begin{figure}[!htb]
  \includegraphics[width=\textwidth]{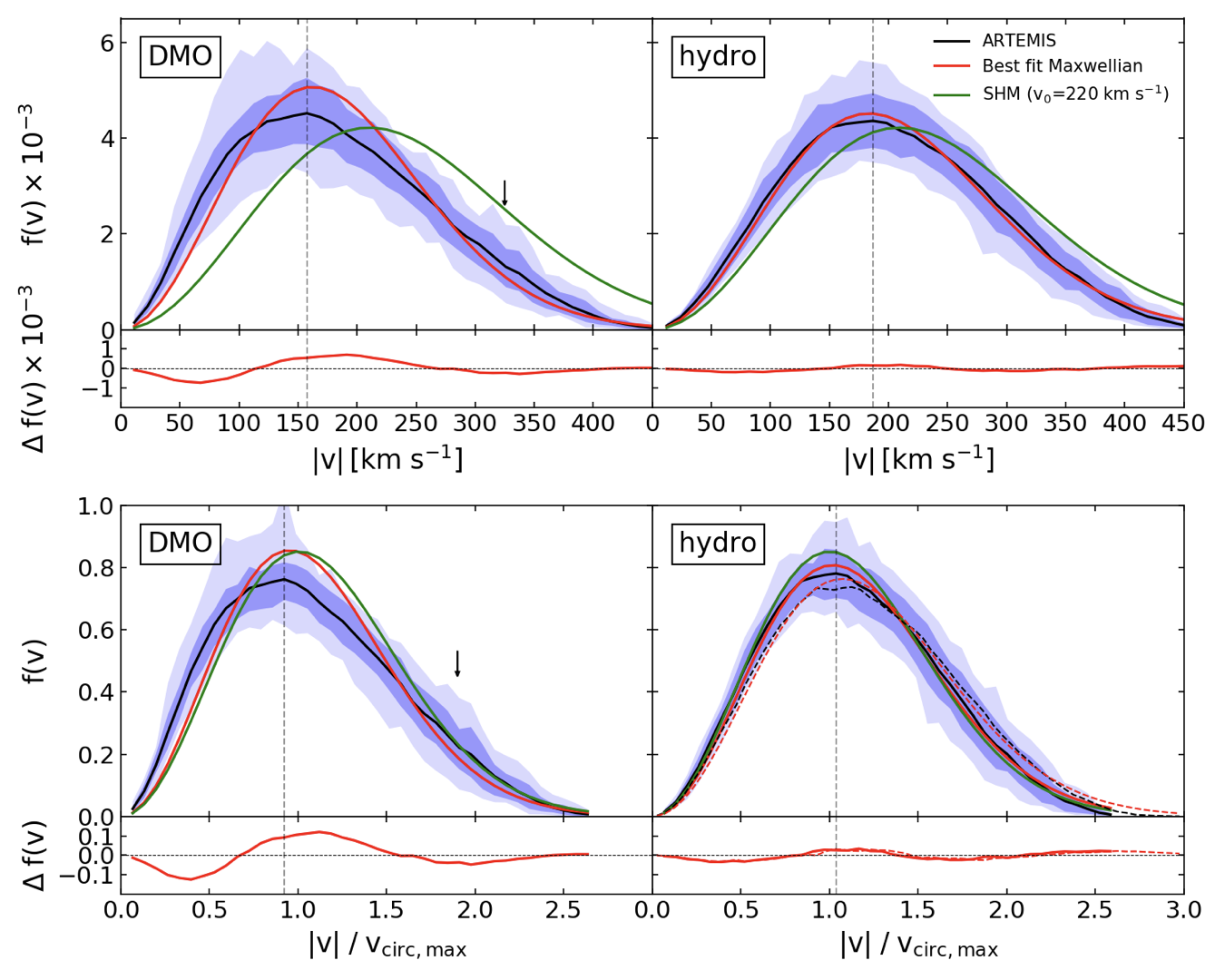}
  \caption{\textit{Top:} DM velocity modulus distributions in the rest frame of the galaxy. The solid black curve represents the median of the velocity modulus distributions for the DMO (left) and hydro (right) simulations. The solid red curve represents the median of the best fitting Maxwellian distributions. The dark and light blue contours enclose 68\% and 95\% of the velocity distribution from all halos. The solid green line represents the velocity distribution determined by the Standard Halo Model with a peak velocity of $\rm{v_0} = 220$ km s$^{-1}$. The lower panels show the residuals between the median distribution of the halos and the median Maxwellian fit. The black arrow points to the high-velocity feature discussed in Section~\ref{sec:SubStr}. \textit{Bottom:} Same as above, but for the DM velocity modulus distributions in the rest frame of each galaxy, normalised by their respective maximum rotational velocities, $v_{\rm{circ,max}}$. The dashed black curve shows the median velocity modulus distribution for the hydro halos normalised by $v_{\rm{circ,max}}$ of the matched DMO halos. The red dashed curve is the median of the best fitting Maxwellian distributions for the DMO normalised velocity distributions.}
  \label{fig:vmod}
\end{figure}

Fig.~\ref{fig:vmod} shows the local DM velocity distributions in our simulations (DMO in the left panel and hydro in the right). The solid black curves show the medians of the local DM $f(\vert v \vert)$ while the dark and light blue contours enclose 68\% and 95\% of the velocity distributions for all the halos.  Owing to the high resolution of our simulations, the simulated solar neighbourhood regions contain a relatively large number of DM particles. This allows us to use small bins in velocity\footnote{We have experimented with various bin sizes and found that the overall shapes of the distribution functions do not change significantly.  For significantly smaller bin widths, though, the data become noisier, while for much large values the occasional distinct features in the distribution can be washed out.}, of $10$~km s$^{-1}$. 

To investigate how well the velocity distributions are fitted by a Maxwellian function, we fit all individual distributions with this function, by allowing the peak velocity, $v_0$, to be a free parameter. The medians of the best fits for both simulation sets are shown in Fig.~\ref{fig:vmod} with red solid curves and the peak of each best fit is indicated by a dashed line. For comparison, we also show the velocity distribution corresponding to the SHM model (green lines) with $v_0 = 220$ km/s. The lower sub-panels show the residuals between the median local DM distribution from the simulations and the median best-fitting functions.  (In Fig.~\ref{fig:App_VDF} in Appendix \ref{sec:appendix} we show the individual velocity distribution functions of each halo.)

Overall, we find that the local DM velocity distributions in the DMO simulations are poorly described by a Maxwellian, with significant differences from this function being seen across the whole velocity range. In contrast, the local distributions in the hydro simulations are relatively well described by a Maxwellian, although slight discrepancies are usually seen near the peaks (Table \ref{tab:chi2} in Appendix \ref{sec:appendix} shows the reduced chi-squared values of the Maxwellian fit for all halos in the DMO and hydro cases).  We will explore the impact of deviations from a Maxwellian distribution on DM direct detection limits in Section~\ref{sec:cross-section}. The lower panels of Fig.~\ref{fig:vmod} show the same velocity distribution as described above but now normalised by the maximum circular velocity of the halos, $v_{\rm{circ,max}}$. Normalising by $v_{\rm{circ,max}}$ removes the mass dependence on the halos velocity distribution, narrowing the overall distribution, although the same conclusions as above can still be made.

In both the DMO and hydro simulations, we observe considerable halo-to-halo variation, with the largest variation seen around the peaks of the distributions. Additionally, the velocity distributions contain stochastic components that are more prevalent at high velocities. These are seen in both sets of simulations, but they are more prominent in the DMO case due to the fact that substructures survive longer in the absence of a massive stellar disc (as discussed below). The averaged DMO velocity distribution also shows a noticeable `bump' at the high-velocity tail, specifically at $v \sim 300$~km s$^{-1}$. This feature will be explored further in Section~\ref{sec:SubStr}, in the context of substructure.

The majority of our halos in both the hydro and DMO simulations have peak velocities less than the value assumed in SHM, i.e. $v_0< v_{0,{\rm SHM}} = 220$~km s$^{-1}$. This is because, as discussed in Section \ref{sec:sims}, both the virial masses and the stellar masses of our simulated halos are somewhat on the lower mass ends of the accepted ranges for the Milky Way. (Also, the Milky Way has a higher stellar mass than typical galaxies in its halo mass range, suggesting that the impact of adiabatic contraction on the density and velocity may be somewhat larger in the Milky Way than typical for this halo mass.)  Specifically, the median peak in the local velocity distribution for DMO simulations is 161.4 km s$^{-1}$, with a full range between $116.8 - 249.8$ km s$^{-1}$, while for the hydro simulations the median peak is at 181.8 km s$^{-1}$, with a range between $152.1 - 237.5$ km s$^{-1}$. The general trend of the increase in $v_0$ from DMO to hydro simulations can be understood in terms of baryons deepening the potential wells of the halos, thus causing the particles to move at higher speeds.

\begin{figure}[!htbp]
  \centering
  \includegraphics[width=\textwidth]{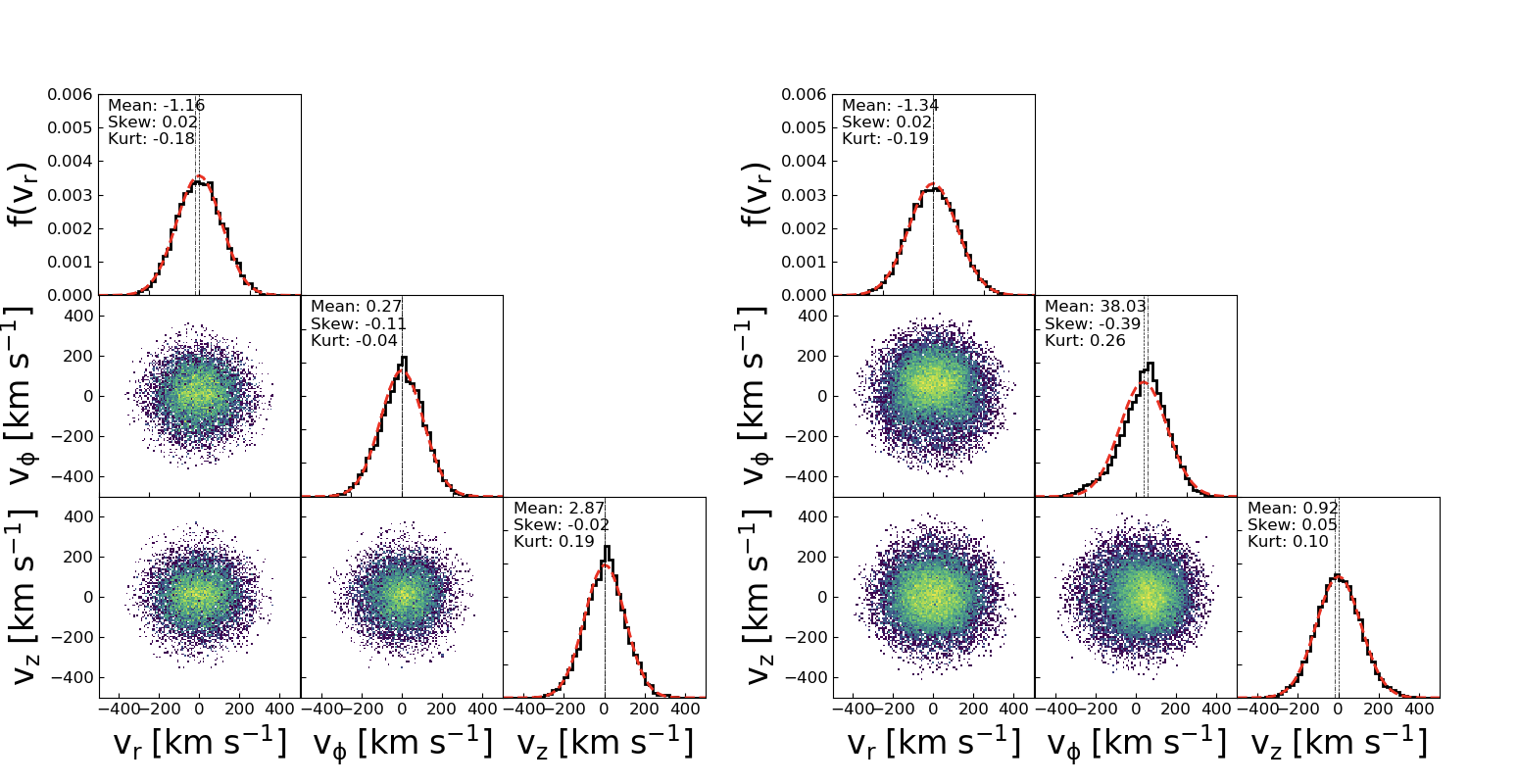}
  \caption{The distribution of velocities in the three components, $(r,\phi,z)$  for the solar neighbourhood of a typical halo (G38) in the DMO simulation. The black line histograms show the individual velocity components, $f(v_r)$, $f(v_\phi)$, and $f(v_z)$, and the red curves show the corresponding best-fit Gaussian functions. Alongside, we also plot the 2D velocity distributions (coloured by the density of the data points). The distribution statistics (mean, skewness and kurtosis) are shown in the upper-left of the plots. \textit{Right:} Same as in the left panel, but now for the hydro simulation.}
  \label{fig:Vrpz_38}
\end{figure}

We also investigate the local distribution of DM velocities along the three cylindrical components, $v_r$, $v_{\phi}$ and $v_z$.  Fig.~\ref{fig:Vrpz_38} shows an example of the local velocity distribution components for one halo (G38), in both the DMO and hydro simulations. The local velocity distribution components are well fitted by Gaussian functions (red curves). Also, we find that the majority of DMO halos show similar distributions in their three velocity components. In the DMO simulations, the means of the $v_r$, $v_{\phi}$ and $v_z$ components are all close to zero. However, in the hydro simulations, the means of the $v_{\phi}$ components show, occasionally, small positive values, indicating net rotation and the presence of `dark discs'.  We will investigate this in more detail in Section~\ref{sec:dark_discs}.

Finally, we infer the escape velocities, $v_{\rm esc}$, for the simulated local velocity distributions. Specifically, we calculate $v_{\rm esc}$ from the high-velocity tail of local halo star particles by following methods outlined by \cite{2014Piffl}, and using:

\begin{equation}\label{vesc}
f(v \, \vert \, v_{\rm esc}, k) \propto (v_{\rm esc} -  v)^k ,
\end{equation}

\noindent for $v < v_{\rm esc}$ and $k$ a parameter constrained by \cite{2014Piffl} to be between $2.3 \leq k \leq 3.7$ from their set of cosmological simulations. Therefore, for our fits, we set $k = 3$ and allow $v_{\rm esc}$ to vary. The median $v_{\rm esc}$ for all our halos is $521.6$ km s$^{-1}$ and the full range is between $509.9 - 631.9$ km s$^{-1}$. This is in good agreement with recent observational measurements from RAVE and, more recently, from \textit{Gaia}. 

In Section~\ref{sec:uncertainties} we will investigate how the WIMP cross-sections calculated within the SHM formalism depend on various assumed values for $\rho_0$, $v_0$ and $v_{\rm esc}$, and in Section~\ref{lz_xenon1T_SHM} we will incorporate in the calculations the full range of these values obtained in the simulations.  

\subsection{Impact of substructure}
\label{sec:SubStr}

\begin{figure}[!htbp]
  \includegraphics[width=\textwidth]{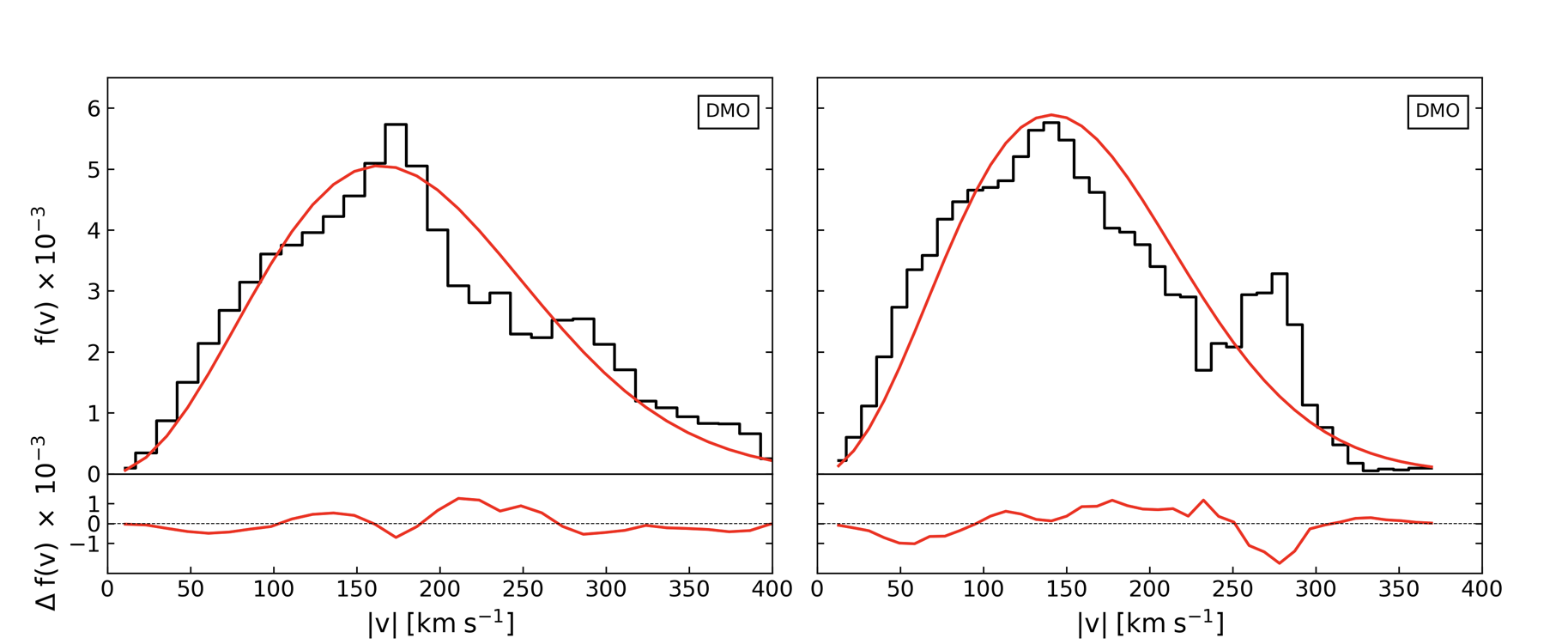}
  \caption{{\it Left panel:} DM $f(\vert v \vert)$ distribution in the rest frame of the G2 halo (with solid black histogram), which shows evidence of substructure in the solar neighbourhood (DMO simulations). The solid red curve represents the best fitting Maxwellian distribution. {\it Right panel:} shows the same distribution, now for halo G28. The lower panels show the residuals between $f(|v|)$ and the best fit Maxwellian function.}
  \label{fig:VDF_sub}
\end{figure}

As mentioned in Section~\ref{sec:localdens_vel}, the average DMO local velocity distribution (see Fig.~\ref{fig:vmod}) contains a peculiar feature at the high-velocity end. This feature can be seen in the 95\% contour and can be attributed to at least two halos in our sample (G2 and G28). The velocity distributions of the solar neighbourhoods in these two systems are shown in Fig.~\ref{fig:VDF_sub}. A prominent peak is seen in the high-velocity tail of each distribution (more so in G28), at $250 < |v| < 350$ km s$^{-1}$, caused by the presence of DM substructure. Also, both distributions clearly deviate from a Maxwellian. 

\begin{figure}[!htbp]
  \centering
  \includegraphics[width=\textwidth]{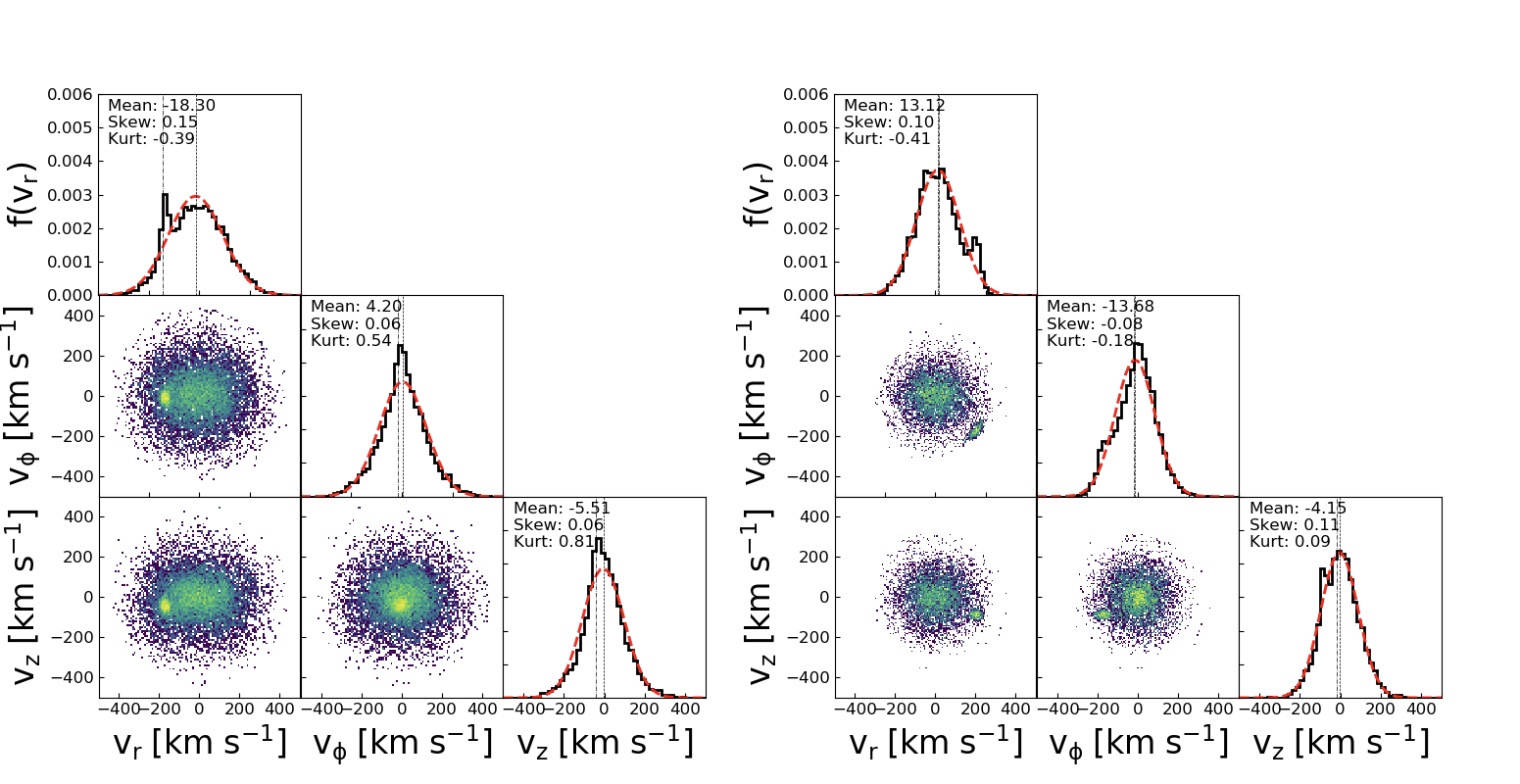}
  \caption{\textit{Left:} The distribution of velocities in the three components, $(r,\phi,z)$  for the solar neighbourhood of the G2 halo in the DMO simulation. The black line histograms show the individual velocity components, $f(v_r)$, $f(v_\phi)$, and $f(v_z)$, and the red curves show the corresponding best-fit Gaussian functions. Alongside, we also plot the 2D velocity distributions (coloured by the density of the data points). The substructure in the local region is clearly seen as the peak in the $v_r$ velocity distribution, as well as the 2D velocity plots. The distribution statistics (mean, skewness and kurtosis) are shown in the upper-left of the plots. \textit{Right:} Same as in the left panel, but for the solar neighbourhood region of the G28 halo, in the DMO simulation. The substructure in the local region is clearly seen as peaks in the $f(v_r)$, $f(v_{\phi})$ and $f(v_z)$ distributions, as well as high density regions in the 2D velocity plots.
}
  \label{fig:Vrpz_Sub}
\end{figure}

We investigate the velocity distributions for these two halos (again, in the DMO simulations) in more detail in Fig.~\ref{fig:Vrpz_Sub}, where we plot the distribution of the $v_r$, $v_{\phi}$ and $v_z$ components in the respective solar neighbourhoods. This shows that the overlap of the secondary peaks in the $f(\vert v \vert)$ was only coincidental. The substructure in the G2 halo has a retrograde $v_r$, peaking at $v_r \approx -150$ km s$^{-1}$, whereas in G28 it peaks at $v_r \approx 200$ km s$^{-1}$. The density plots in velocity space show small clusters at $(v_r,v_{\phi},v_z) \approx (-150,0,-50$) km s$^{-1}$ for G2 and at $(v_r,v_{\phi},v_z) \approx (200,-150,-75$) km s$^{-1}$ for G28. The 2D velocity plots indicate that local halos are anisotropic (see also the 2D plots in the Fig.~\ref{fig:Vrpz_38}). Deviations from gaussianity are seen in all the three components of the local velocity distributions for G28, near the location of the cluster. G2 shows a similar deviation at $v_r \approx -150$ km s$^{-1}$. 

Interestingly, we observe no separate velocity peaks in the hydro versions of the G2 and G28 systems, in their corresponding solar neighbourhoods. This suggests that some of the substructure seen in the DMO simulations may be erased in the presence of baryons. Previous hydrodynamical simulations have found that the number of DM subhalos in the inner regions can decrease by about $50\%$ compared to DMO simulations, due to additional tidal disruption induced by the stellar galactic disc (e.g., \cite{2016Sawala,stafford2020}). We find a similar result in our simulations (not shown here quantitatively, however the paucity of subhalos in the inner regions of galaxies is immediately apparent in comparing the left and middle columns of Fig.~\ref{fig:halos}). This also suggests that using DMO simulations may occasionally overestimate the sensitivity to DM due to the relatively long-lived nature of substructure in DMO simulations compared to that in hydro simulations.  

In principle, any DM substructure in the solar neighborhood, either in the form of bound clumps or tidal streams, may lead to a boost in the DM signal, e.g. by inducing a step-like feature in the energy recoil spectrum \cite{2005Freese, 2001Gelmini}, and thus affect the direct DM detection limits. The strength of the signal depends not only on the mass in the DM substructure, but also on the direction of motion of the DM particles relative to the Earth \cite{2010Kuhlen}. Several tidal streams are known to pass through the solar neighbourhood. In addition to the Sagittarius stream,  \textit{Gaia} has revealed several other substructures  \citep{2016GAIAA&A...595A...2G, 2018GAIAA&A...616A...1G}. One of them is a tidal debris from a massive satellite galaxy that fell in $\sim 10$~Gyr ago, dubbed the \textit{Gaia} sausage \citep{2018Natur.563...85H}. Other known streams include Nyx \citep{2019arXiv190707190N, 2019arXiv190707681N} or the S1 and S2 streams \citep{2018Myeong}. The Sagittarius stream is likely to have a non-negligible contribution to the local DM distribution \cite{2012Purcell}, while \textit{Gaia} sausage is expected to have a modest effect on the DM detection rates \cite{2019Evans,2020JCAP...07..036B}. The S1 stream can lead to an increase in the number of high energy nuclear recoils and a slight improvement of DM detection rates \cite{OHare2018DarkDetectors}, particularly for directional experiments since this stream is retrograde. The S2 stream can also lead to multiple  effects in the DM signal \citep{O'Hare2020}. 

Although our simulations are not suited to model the observed streams in the Milky Way specifically, they do include the contribution of local DM substructures and so we can gauge, in a broader sense, the effect that this type of features may have on the direct detection limits (see Section~\ref{lz_xenon1T_SHM}). Generally, we find that substructures that are massive enough to increase the local DM detectability rates are not very common in our simulations, particularly in the hydro simulations where such features are efficiently erased by tidal forces.

\section{Larger scale changes to the DM structure}
\label{sec:non-local}

We have seen that the inclusion of baryons generally results in an increase in both the local (solar neighbourhood) density and velocities of DM particles; a result of adiabatic contraction of the DM halos in response to the baryons.  Here we explore further how this is achieved in practice, by examining the impact on the halo shape and the prevalence of dark discs.

\subsection{DM halo shapes}
\label{sec:shapes}

Here we contrast the shapes of the simulated DM halos, between the DMO and hydro simulations. We determine the shapes of the simulated DM halos from the ratios of the principal axes, $a$ (major), $b$ (intermediate) and $c$ (minor), which are calculated from the eigenvalues of the mass distribution tensor within an inner region of radius $30$~kpc. The principal axes are used to calculate the DM halo sphericity, $S=c/a$ and DM halo triaxiality, $T=\frac{a^2-b^2}{a^2-c^2}$. 

\begin{figure}[!ht]
  \includegraphics[width=\textwidth]{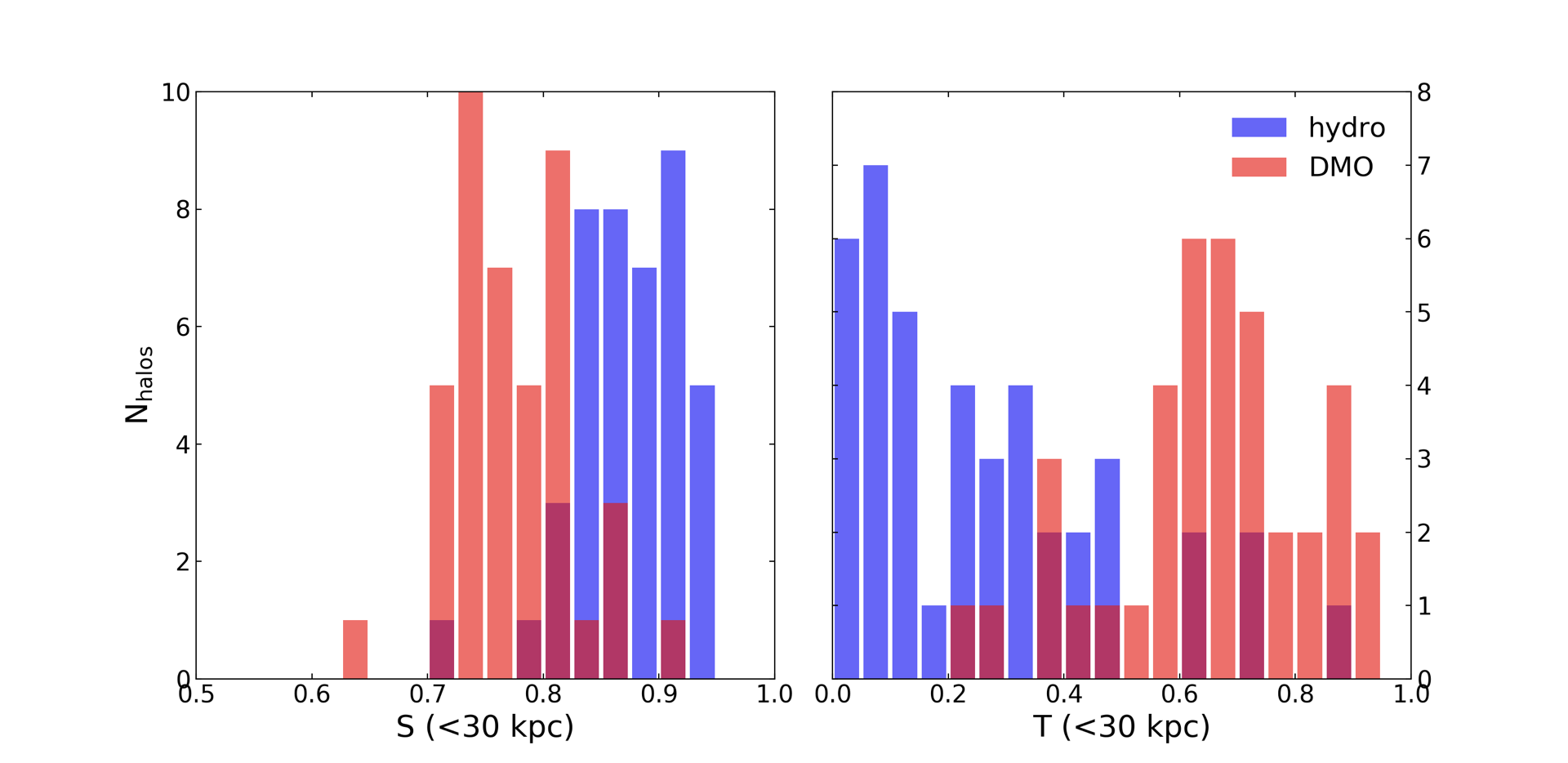}
  \caption{\textit{Left}: Distribution of sphericity, $S$, of all Milky Way-mass halos in the hydro (blue) and DMO (red) cases, calculated for regions within $30$~kpc of the galactic centre. \textit{Right}: The distribution of the triaxiality, $T$, for all Milky Way-mass halos in the hydro and DMO cases, also calculated within $30$~kpc. Baryonic processes lead to more spherical and less triaxial halos.}
  \label{fig:S_T}
\end{figure}

Fig.~\ref{fig:S_T} shows the sphericity and triaxiality parameters for simulated halos used in this work for both the DMO and hydro cases. The hydro halos tend to be much more spherical ($S \rightarrow 1$) than the DMO halos, which is a well-known effect of the inclusion of baryons \cite{Kazantzidis2004TheHalos, Debattista2007TheSubstructures, Tissera2010DarkFormation, Kazantzidis2010TheDisks}. This result is reassuring, as recent analysis of stellar halo kinematics with \textit{Gaia} DR2 suggests that the DM halo of the Milky Way is, at least in the inner region, nearly spherical \cite{2019Wegg}. The hydro halos also tend to be more oblate ($T \rightarrow 0$), in contrast to the prolate ($T \rightarrow 1$) distribution of the DMO halos. This is also expected, as the gas infall and the associated formation of a large-scale disc component can result in significant changes in the halo shapes \cite{Gunn1991DissipationalGasdynamics, Dubinski2002TheHalos, Debattista2007TheSubstructures}.   

\subsection{Probability of hosting a dark disc}
\label{sec:dark_discs}

One avenue for the formation of a dark disc is through the accretion of dwarf satellite galaxies and their subsequent tidal disruption in the Galaxy. The accreted dark disc is likely to be formed from tidal debris from satellites incoming on low-inclination orbits \cite{2008Read}. A stellar disc also drags the incoming satellites towards the disc plane where they are more easily torn apart by tides \cite{Read2009AGalaxies}. Alternatively (or in addition to), adiabatic contraction due to the stellar disc may also lead to the formation of a dark disc. The presence of a dark disc may have implications for direct detection of DM, as a low-velocity dark disc with respect to the Earth can increase the rate of detection at lower recoil energies \cite{2009Bruch, Read2009AGalaxies}.

\begin{figure}[!htbp]
  \includegraphics[width=\textwidth]{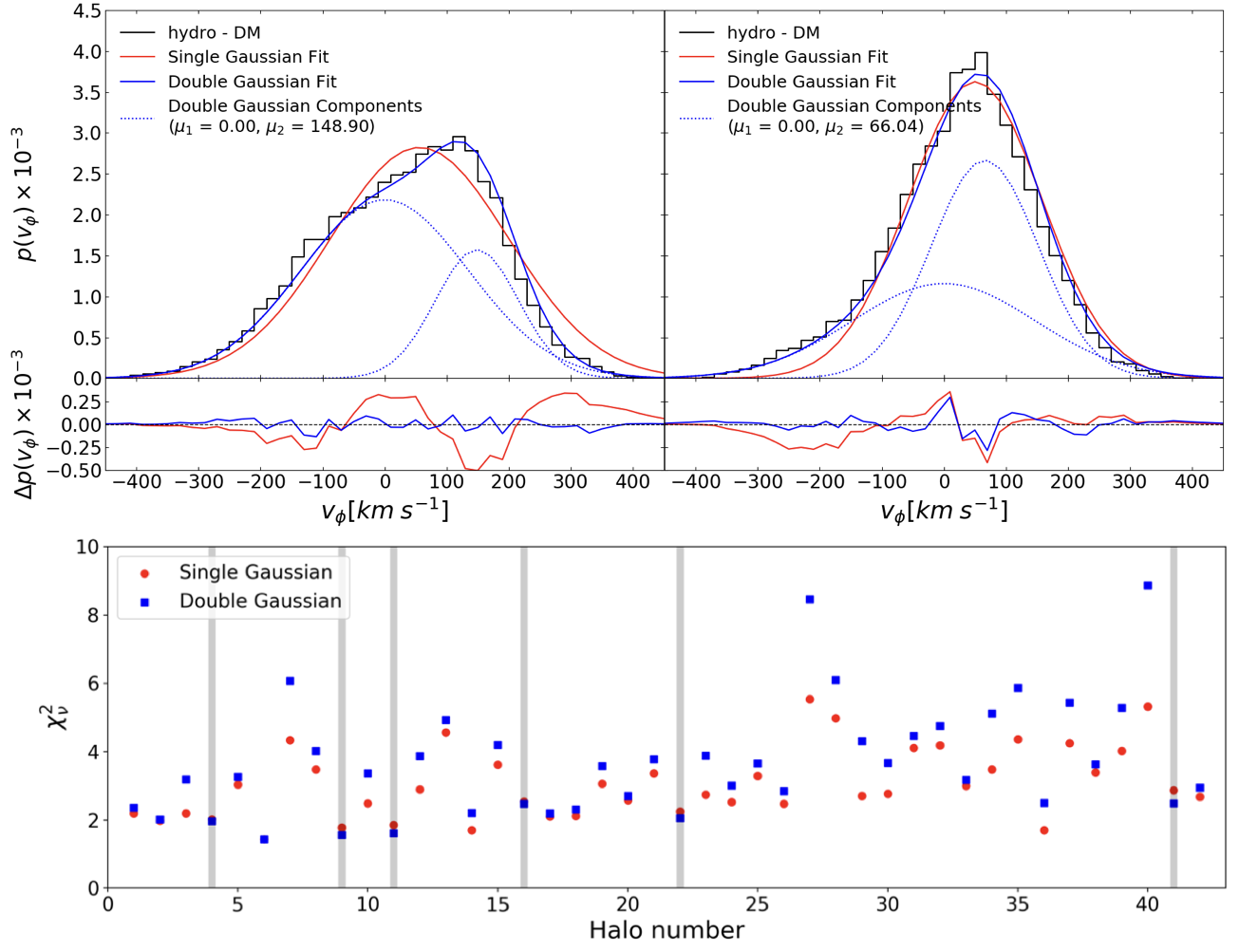}
  \caption{\textit{Upper panels:} The azimuthal velocity DM distribution, $v_{\phi}$ for the G11 halo (left), which is the halo with the most prominent dark disc component in our sample, and for the G38 halo (right) in the hydro simulations. The black histogram shows the DM distribution from the simulations; the solid red curve shows the best-fitting Gaussian to the distribution; the solid blue curve shows the best-fitting double Gaussian, with the separate components shown as blue dashed curves. The difference between the fits is shown below the distributions. \textit{Lower panel}: The reduced $\chi^2$ values for all halos in the hydro simulations, comparing the goodness of fit between the single Gaussian fit and the $v_{\phi}$ DM distributions (red circles) and between double Gaussian fits and the $v_{\phi}$ DM distribution (blue squares), respectively. Most halos are better fitted by a single Gaussian. The six halos that are better fitted by a double Gaussian are highlighted with grey bands.}
  \label{fig:Schaller}
\end{figure}

To quantify how often dark discs occur in the \texttt{ARTEMIS} simulations, we first use the methodology of \cite{Schaller2016TheGalaxies, 2010Ling}, which identifies a dark disc when a double Gaussian better fits the local DM $v_\phi$ distribution than by a single Gaussian. The upper panels of Fig.~\ref{fig:Schaller} illustrate the $v_\phi$ distribution of DM in the solar neighbourhoods of two halos, G11 and G38  (black lines), with the best fits for single and double Gaussians shown with the solid red and blue lines, respectively. The components of the double Gaussian are also shown (with blue dashed lines). For the fitting, one Gaussian component is fixed at $v_{\phi} = 0$ km s$^{-1}$ (corresponding to the assumption of a non-rotating DM halo), while the second is allowed to vary freely. The difference between the two fits and the $v_{\phi}$ distribution is shown below. 

Halo G11 provides the strongest evidence for the existence for a dark disc among all halos in our sample. Its second component of the double Gaussian exhibits a large $v_{\phi}$ value centred around $148.9$ km s$^{-1}$, indicating a significant prograde motion. In contrast, halo G38 prefers a single Gaussian fit, with the second component centred around $v_\phi \approx 66.0$ km s$^{-1}$. Not unexpectedly, we find that dark discs are more prevalent in systems in which $f(\vert v \vert)$ deviates more strongly from a single Gaussian (or Maxwellian). Interestingly, all dark disc components are found to be co-rotating with the stellar disc.

The lower panel of Fig.~\ref{fig:Schaller} shows the prevalence of dark discs in the entire sample. The reduced $\chi^2$ values for the single Gaussians are shown with red circles and those for double Gaussians with blue squares. For clarity, the grey columns highlight the halos that are better fitted by a double Gaussian. We find that in the majority of our halos, the $v_\phi$ distribution is better fitted by a single Gaussian. By a conservative estimate, this method retrieves that $\approx 14 \%$ of halos (6 out of the 42 in our sample) contain a dark disc. Previous work investigating the existence of dark discs in Milky Way analogues in \texttt{EAGLE} found, with the same method, that only 1 out of 24 halos has a detectable dark disc \citep{Schaller2016TheGalaxies}. One possible explanation for why we find more dark discs in \texttt{ARTEMIS} than found in \texttt{EAGLE} is due to the higher stellar mass fractions in \texttt{ARTEMIS}, which are in better agreement with observations \cite{2020Font} and which lead to enhanced adiabiatic contraction \citep{stafford2020}.

\begin{figure}[!htbp]
  \includegraphics[width=\textwidth]{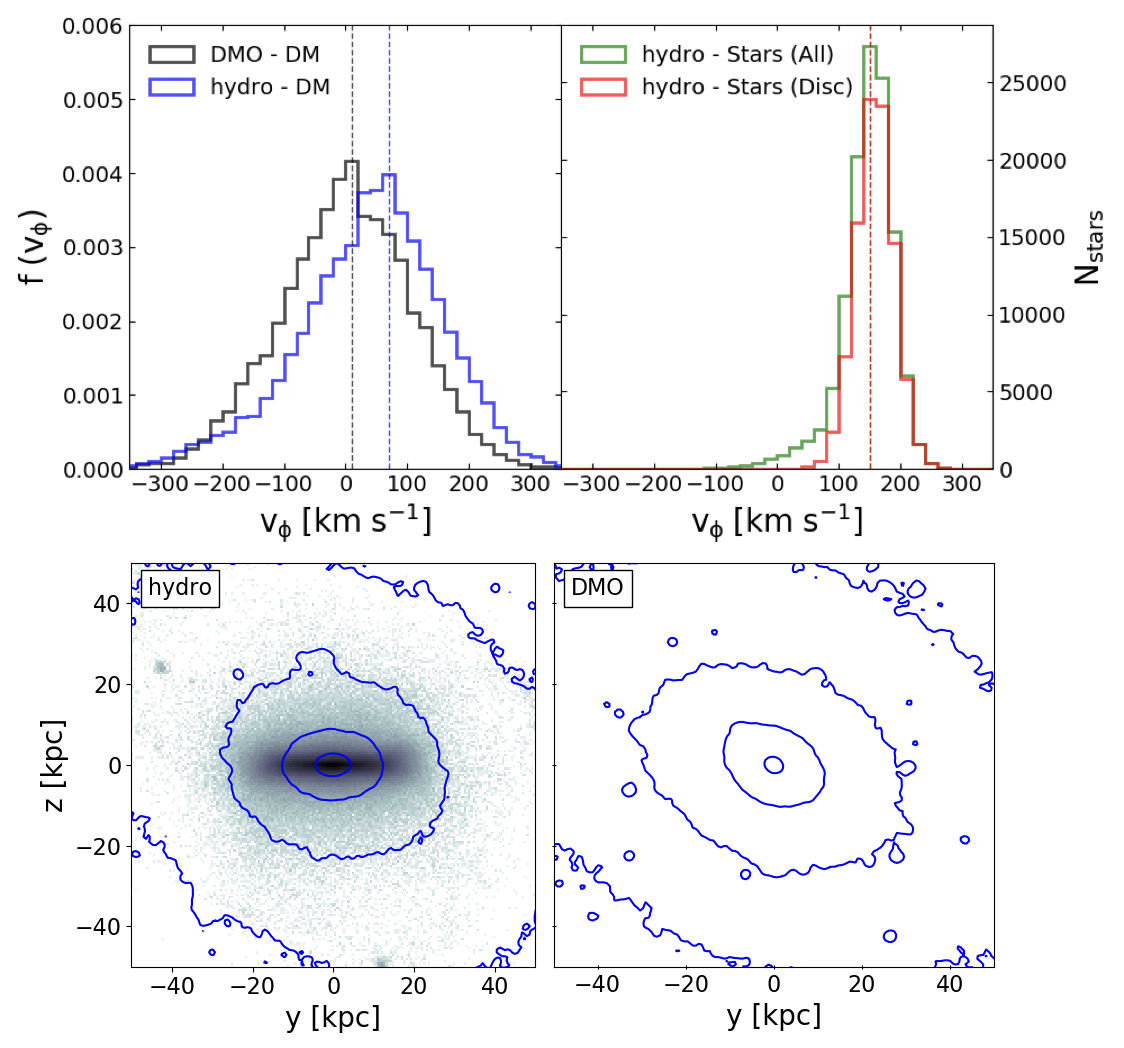}
  \caption{\textit{Top row}: The local distribution of rotational velocities for the G38 halo. \textit{Left:} The DM distributions for the DMO and hydro simulations are shown in black and blue, respectively.  The dashed lines in their respective colours show the peaks of the distributions. \textit{Right:} The stellar distributions for all stars and disc stars are shown in green and red, respectively. The dashed lines represent the peak velocity in the distributions. {\textit{Bottow row.} \textit{Left}: The projected DM density contours for G38 in the hydro simulation. The underlying points show the distribution of the stellar component, with the disc clearly visible at the centre. \textit{Right}: The projected DM density contours in the G38 system in the DMO simulation.}}
  \label{fig:Vphi_58}
\end{figure}

Fig.~\ref{fig:Schaller} also shows that, in some cases where the double Gaussian fit is preferred, the difference in $\chi^2$ from a simple Gaussian is not sufficiently large to be conclusive. Nevertheless, these (and other) halos clearly exhibit non-Maxwellian velocity distributions and display significant net rotation. This suggests that the above method may not capture the existence of dark discs accurately enough. 

Fig.~\ref{fig:Vphi_58} exemplifies this point with the G38 halo, which is better fitted by a simple Gaussian, yet it exhibits other indications that it contains a dark disc, for example in its DM halo shape and rotation characteristics. The top left panel shows the azimuthal velocity $v_\phi$ distribution of DM in the solar neighbourhood of this system, both in the DMO and the hydro simulations. In the hydro case, the peak of the DM distribution is skewed towards that of the stars. The local stellar $v_{\phi}$ distributions are shown in the top right panel, for both all stars (green) and the disc (red). The local DM component co-rotates with the stellar disc, with a peak $v_{\phi} \approx 75$ km s$^{-1}$, which is roughly half of that of the stellar disc in this galaxy. The bottom row in Fig.~\ref{fig:Vphi_58} shows, with contour lines, the DM distribution in this galaxy in the presence and absence of baryonic effects. The introduction of baryons into the simulations causes the central region of the DM halo to become oblate and aligned with the stellar disc, the latter being shown as a background stellar particle density (for a similar result, see also \cite{Read2014TheDensity}). Taken together, the oblate shape of the DM component and its prograde rotation suggest that this system contains a dark disc. Note however that the evidence indicates that the local DM rotates as a whole (which is also supported by the fact that the $f(|v|)$ prefers a single Gaussian rather than two). This suggests that, in this case, the DM halo has acquired its rotation due to the presence of baryons. These can cool and form a rotationally-supported disc, thus causing a shift in the velocity distribution of DM.

\begin{figure}[!htbp]
  \includegraphics[width=\textwidth]{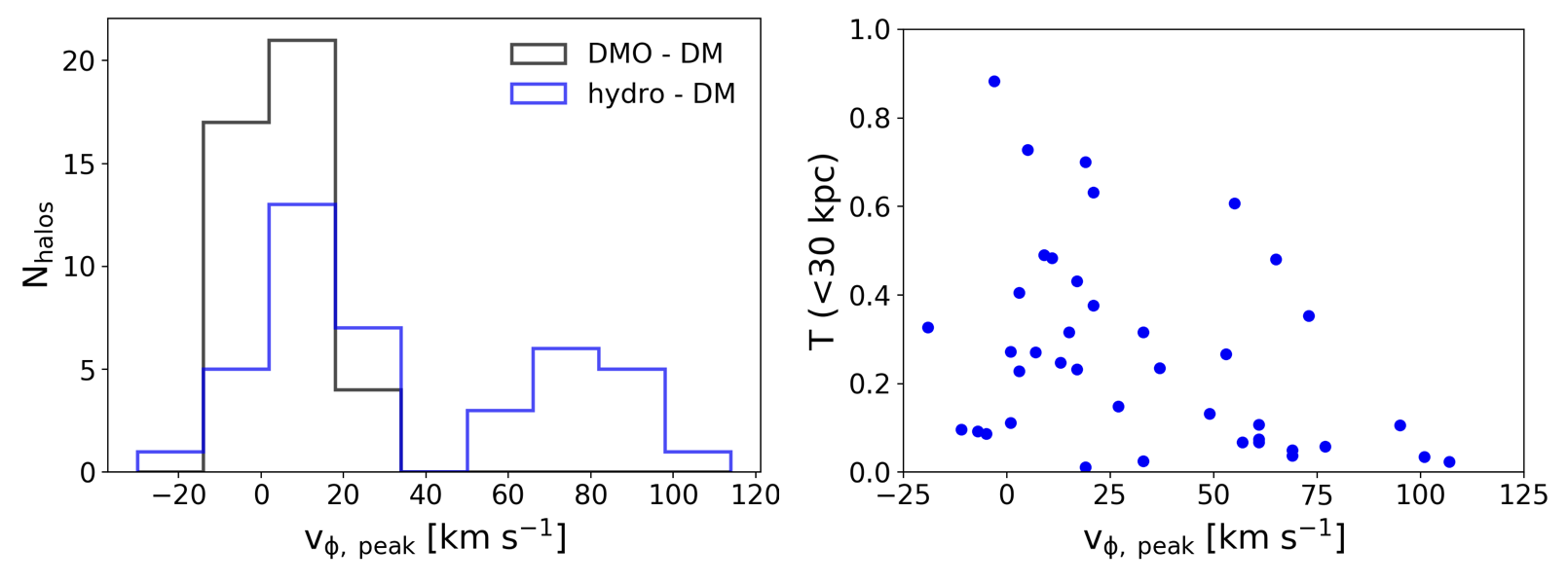}
  \caption{\textit{Left:} Histograms of the peak azimuthal velocity DM distribution $v_{\phi}$ for the hydro (blue) and DMO (black) \texttt{ARTEMIS} halos. \textit{Right:} The peak azimuthal velocity, $v_{\phi}$ versus triaxiality $T$ within $30$~kpc of the galactic centre for the hydro halos.}
  \label{fig:Vphi_peak}
\end{figure}

In order to identify other systems which are best fitted by single Gaussians, yet display dark disc characteristics, we further investigate the distribution of the peak azimuthal velocities in all halos. This is shown in the left panel of Fig.~\ref{fig:Vphi_peak}, for both the hydro and DMO simulations. In the hydro simulations, there is a distinctive category of halos ($15$ in total) with significant prograde rotation of their local DM component ($v_{\phi} > 50$ km s$^{-1}$; the peak $v_{\phi}$ for DM and stars for all halos are shown in Table \ref{tab:halo_properties}). All $6$ halos found previously to contain dark discs via the double Gaussian fitting method are in this category.  Note that, if one defines a dark disc as a separate rotating component with an overall non-rotating halo, then the other $10$ halos would not, strictly speaking, qualify as dark discs.  If, however, one also includes rotating, flattened DM halos (i.e., the whole inner halo is rotating, rather than two separate components) that are aligned with the stellar disc, then our work suggestions that the fraction of Milky Way-mass galaxies with dark discs can be as high as $\approx 36\%$.

There is a marked difference in the rotation pattern of local DM components in the hydro versus DMO simulations in the sense that net prograde motions are conspicuously missing in the DMO simulations. Clearly, the baryonic effects (namely the adiabatic contraction referred to earlier) play a role in the emergence of dark discs. The right panel of the same figure shows a (mild) anti-correlation between the triaxiality parameter $T$ of galaxy systems in the hydro simulations and their peak $v_{\phi}$. This suggests that systems that have the fastest prograde motion of their local DM components also tend to be more oblate. 

As an additional test to quantify the importance of dark discs, we focus on the density enhancement that results.  Specifically, we compute the following two metrics: i) the ratio of our fiducial cylindrical DM density estimate to one where the DM density is estimated using a spherical shell of the same radius and width as the cylinder, both from the hydro simulations; and ii) the ratio of the fiducial cylindrical DM density from the hydro simulations to a spherical shell-based estimate from the corresponding DMO halo.  Note that the first metric will somewhat underestimate the importance of a dark disc, since the dark disc will also contribute to the spherical shell estimate.  On the other hand, the second metric will somewhat overestimate the importance of a dark disc, since the DM density in the hydro simulation would be increased by normal (spherical) adiabatic contraction.

For the first metric we compute a mean ratio of 1.27 with a standard deviation (intrinsic scatter) of $\pm 0.20$.  For the second metric we compute a mean ratio of 1.69 with a standard deviation of $\pm0.34$.  Thus, the presence of a dark disc increases the solar neighbourhood DM density by tens of percent, constituting a significance enhancement with respect to systems that do not possess such a structure.

\section{Effects on DM direct detection limits}
\label{sec:cross-section}

In the preceding sections, we have shown that there is significant halo-to-halo scatter in the solar neighbourhood DM density and velocity distributions and that baryons also play an important role in setting these quantities.  Here we explore the impacts of the scatter and baryons on the DM direct detection limits using the simplified model outlined in Section \ref{sec:SHM}.

Specifically, we show the effects of incorporating the halo-to-halo scatter in the density and velocity distributions from \texttt{ARTEMIS} on the exclusion limits for the XENON1T and LZ experiments, in the spin-independent WIMP-nucleon cross-section--mass plane.  Before doing so, however, we show the impact of systematically varying each of the SHM parameters on the detection limits, to build some intuition for the simulation-based results.

\subsection{Exploring variations to the SHM}
\label{sec:uncertainties}

Direct detection experiments have set increasingly strong constraints on the cross-section of a WIMP-nucleon interaction under the assumption of the SHM, adopting fixed values for $\rho_0$, $v_0$ and $v_{\rm esc}$. However, as discussed in Section~\ref{sec:SHM}, there are significant variations in the measured values for these parameters.  Also, as shown in Section~\ref{sec:localdens_vel}, the simulations also show variations in these values.
Here we show the effects of varying the astrophysical parameters on the determination of the WIMP-nucleon cross-section limits. For these calculations, we assume a generic Xenon detector with a 1000 kg per year exposure, zero observed events and 100$\%$ efficiency across recoil energies from 5 to 40 keV$_{nr}$.

\begin{figure}[ht!]
  \centering    
  \includegraphics[width=0.7\textwidth]{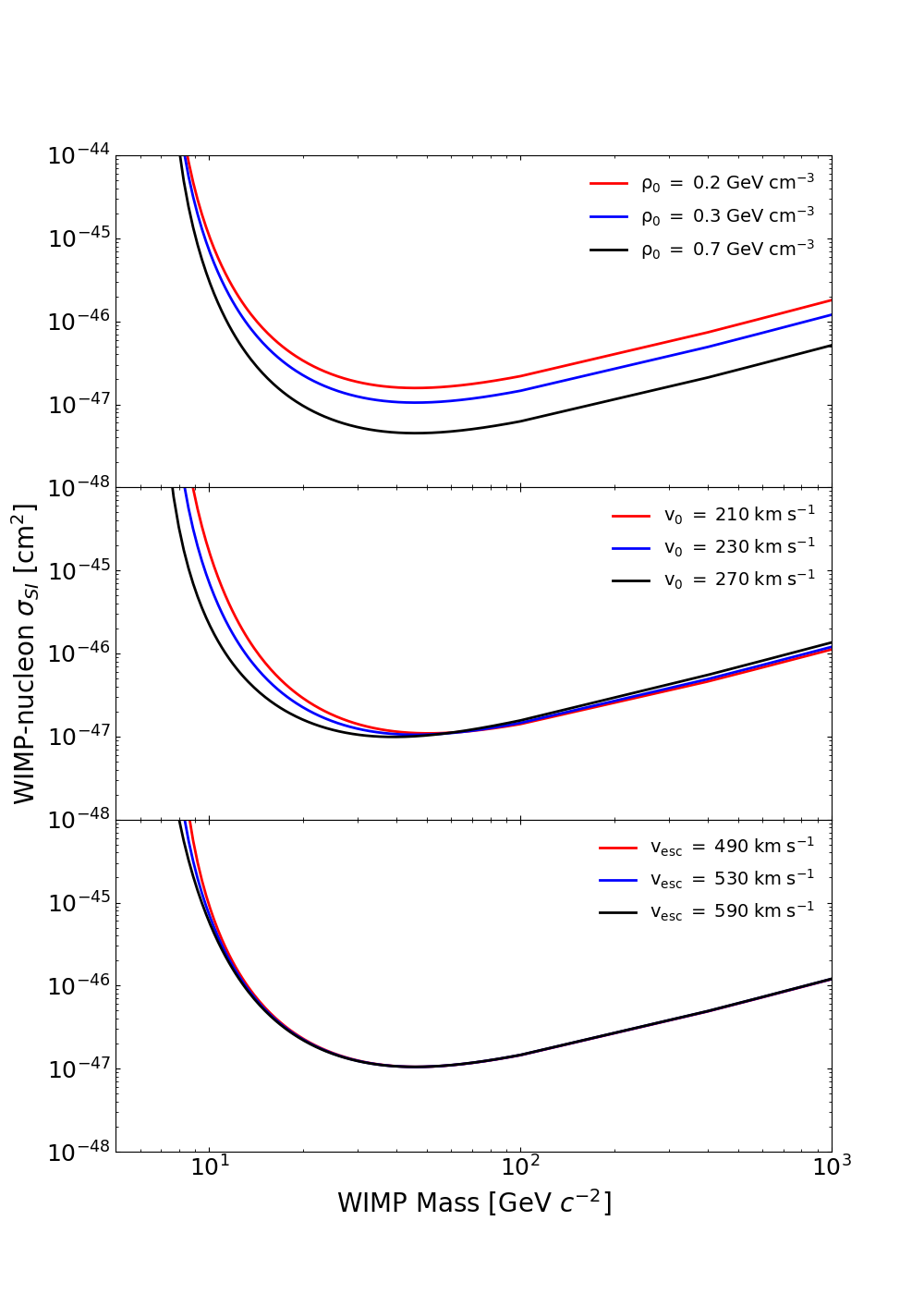}
  \caption{The 90\% confidence level limits for a single halo with a Maxwellian distribution of local DM velocities, assuming SHM and varying $\rho_0$, $v_0$ and $v_{\rm esc}$ independently (i.e. varying one parameter at a time, while keeping the other two fixed). The \textit{upper} panel shows the variations in $\rho_0$: 0.2, 0.3 and 0.7 GeV cm$^{-3}$, respectively; the \textit{middle} panel shows the $v_0$ variations: 210, 230 and 270 km s$^{-1}$; and the \textit{lower} panel shows the variations in $v_{\rm esc}$: 490, 530 and 590 km s$^{-1}$. In each panel, the black lines represent (approximately) the typical values used in SHM models, while the red, blue lines represent (approximately) the minimum and maximum ranges found in observations and/or simulations.}
  \label{fig:SHM}
\end{figure}

The three panels in Fig.~\ref{fig:SHM} show the results of varying the local DM density (upper), local DM peak velocity (middle) and escape velocity (lower), respectively, whilst keeping the other parameters fixed. The range in which each parameter is chosen to cover (roughly) the current uncertainties in the observational measurements (see Section~\ref{sec:SHM}). The areas above the curves indicate the regions that direct detection experiments are sensitive to.

As can be seen from equation \ref{eq:BBRate}, the interaction rate and hence cross-section, directly scales with the local density of DM, in that a larger density gives rise to more scattering events and a stronger limit. Our model shows that the WIMP-nucleon cross-section can vary by half an order of magnitude when considering the plausible range of values for $\rho_0$.

The combination of equations \ref{eq:vmin} and \ref{eq:BBRate} reveals the dependence on the velocity model parameters $v_0$ and $v_{\rm esc}$. This dependence is strongest at low DM masses where $m_{\rm DM}$ $\ll$ $m_{\rm N}$, and the minimum DM velocity required to produce detectable recoil energy is inversely proportional to the DM mass (see equation \ref{eq:vmin}). When $v_{\rm min}$ is comparable to $v_0$ ($m_{\rm DM} < \sim$40~GeV), the changes in $v_0$ strongly affect the velocity integral in equation \ref{eq:BBRate} and, as a result, the expected rate. When $v_{\rm min}$ is close to $v_{\rm esc}$ ($m_{\rm DM} < \sim$10~GeV) then the velocity integral becomes very constrained, and the changes in $v_{\rm esc}$ also strongly affect the expected rate of DM interactions with Xe nuclei.

\subsection{Direct detection limits using \texttt{ARTEMIS}}
\label{lz_xenon1T_SHM}

In the following, we incorporate the local DM densities, local peak velocities (determined from the best-fitting Maxwellian distributions) and escape velocities measured directly from the simulations into the direct detection limits methodology. For this, we use experimental parameters of LZ and XENON1T experiments to better compare with the experimental results. We show results for both the DMO and hydro simulations with the aim of comparing the two and therefore to determine the importance of including baryonic physics in the predicted direct detection limits. 

We note that our method does not incorporate a background model.  However, this is unimportant for our purposes, as we are mostly interested in the \textit{relative} effects of varying the velocity distribution function and DM density, as guided by the simulations. When reproducing the expected limit of a specific detector, we consider the convolved detection and selection efficiencies.  The efficiencies related to the recoil energy are included by interpolating the experimental nuclear recoil efficiencies \cite{LZ_2018,2018Aprile}. To further ease the comparison, the variations between the experimental and simplified models are accounted for by scaling the calculated limit to the 100 GeV $c^{-2}$ experimental result (see Appendix~\ref{sec:appendixLimits}).

As discussed in Section~\ref{sec:SHM}, the limits placed on the WIMP-nucleon cross-section depend not only on several local DM properties, but also on various experimental parameters such as the sensitive material and mass, energy-dependent detection and selection efficiencies, background events, and the number of days the experiment runs for (i.e. live-days). The XENON1T direct detection experiment \cite{2017Aprile, 2018Aprile} provides constraints on value on the WIMP-nucleon cross-section, with an upper limit on the WIMP-nucleon spin-independent cross-section of $4.1\times10^{-47}$ cm$^2$ at a WIMP mass of 30 GeV c$^{-2}$. This experiment has been carried out over 278.8 live-days with a 1.3-tonne detector, equivalent to 1.0 tonne over a year. The most sensitive projections from the LZ experiment were carried out with a 5700 kg Xenon detector over 1000 live-days \citep{LZ_2018}. This places an upper limit on the WIMP-nucleon spin-independent cross-section of $1.4\times10^{-48}$ cm$^2$ at a WIMP mass of $40$~GeV $c^{-2}$. For our calculations, we use the full recoil energy of 1 - 60 $\rm{keV_{nr}}$ covered by both detectors and apply the energy-dependent efficiencies extracted from the publications.

\begin{figure}[tbh!]
  \includegraphics[width=\textwidth]{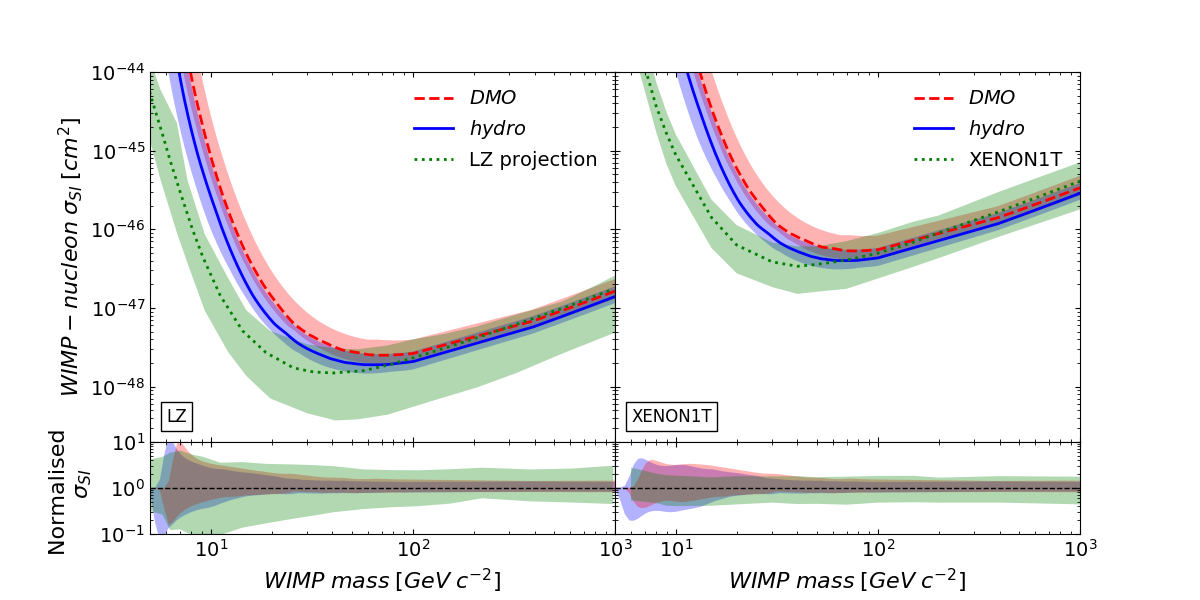}
  \caption{\textit{Left}: The 90\% confidence level limits for the \texttt{ARTEMIS} halos using the LZ parameters \cite{LZ_2018}. The median upper-limits are shown as red dashed and blue solid lines for the DMO and hydro case, respectively.  The contours enclose 68\% of all the individual exclusion limits for all of the halos. The green dotted line shows the published experimental projections for LZ assuming the SHM and the green shaded region corresponds to the 1$\sigma$ experimental uncertainties on the limits. \textit{Right}: Same as left, but now using the XENON1T experimental parameters to calculate the 90\% confidence level limits. The lower panels show the uncertainties on the WIMP cross-sections normalised by their respective 90\% confidence level limits.}
  \label{fig:CS}
\end{figure}

The exclusion limits based on the simulated Milky Way-mass halos, which use the best fitting Maxwellian velocities, and the LZ direct detection experiments can be seen in the left panel Fig.~\ref{fig:CS} (with red for DMO and with blue for hydro).  The coloured regions enclose 68\% of all the individual exclusion limits for the 42 halos, while the solid and dashed lines show the median values for all the halos. The right panel of Fig.~\ref{fig:CS} shows the same result but using the XENON1T experimental parameters. The green dotted lines in the left and right panels indicate the 90\% confidence level limits from LZ and XENON1T collaborations, respectively. These exclusion limits were calculated using the SHM with a peak speed of $220$ km s$^{-1}$, a local DM density of $0.3$ GeV cm$^{-3}$ and an escape velocity of $544$ km s$^{-1}$, which are the assumed values for both XENON1T and LZ experiments.  The green shaded regions correspond to the published 1$\sigma$ experimental uncertainties (projections in the case of LZ) on the published limits. The lower panels in Fig.~\ref{fig:CS} show the uncertainties on the WIMP-nucleon cross-section exclusion limits when normalised by the median 90\% confidence level limits corresponding to the individual cases, in order to more clearly demonstrate how the halo-to-halo variation in the limits compares with the typical experimental uncertainty.

The exclusion limits for the simulations (DMO and hydro) are higher in amplitude at low WIMP masses ($< 50$ GeV $c^{-2}$) compared to the published limits for XENON1T and LZ.  This is just due to the fact that the latter use the SHM with $v_0 = 220$ km/s, whereas the median $v_0$ for the ARTEMIS halos is closer to 180 km/s (see Fig.~\ref{fig:vmod}).  However, our focus is primarily on relative effects, namely the relative effect of including baryons and the halo-to-halo scatter.

Our results show that the hydro halos place lower upper limits on the cross-section compared to their matched DMO counterparts. We also observe a large halo-to-halo variation in the WIMP cross-sections for both hydro and DMO halos. Typically, the $1\sigma$ scatter in the exclusion limits is a factor of $\approx1.5$ but increases to a factor of several at low WIMP masses.  Both of these results can, in part, be explained by considering the local DM densities; the WIMP-nucleon cross-section is inversely proportional to the local DM density, and the densities are higher in the solar neighbourhood in the hydro simulations.  In addition, the enhanced peak velocities in the hydro simulations lower the cross-section constraints at the low mass end.  The halo-to-halo variation the in DM density and peak velocities propagates through to the spread seen in the exclusion limits.

It is important to note that the halo-to-halo variation in the exclusion limits is smaller than the current experimental uncertainties for XENON1T but not by large amounts (compare the width of the green shaded region with the widths of the red and blue shaded regions).  In fact, at low WIMP masses of $< 20$ GeV $c^{-2}$, the simulated scatter is larger than the experimental uncertainties for XENON1T.  The (projected) experimental uncertainties in the LZ experiment are slightly larger than that of XENON1T, however these are still comparable at low WIMP masses, as the modelling uncertainty becomes increasingly important.  Given that this is the case, it suggests that the modelling uncertainties (i.e., in $\rho_0$, $v_0$, $v_{\rm esc}$, $f(v)$) should be included in the overall error budget in order to derive a conservative estimate of the cross-section limits.

\begin{figure}
  \centering    
  \includegraphics[width=\textwidth]{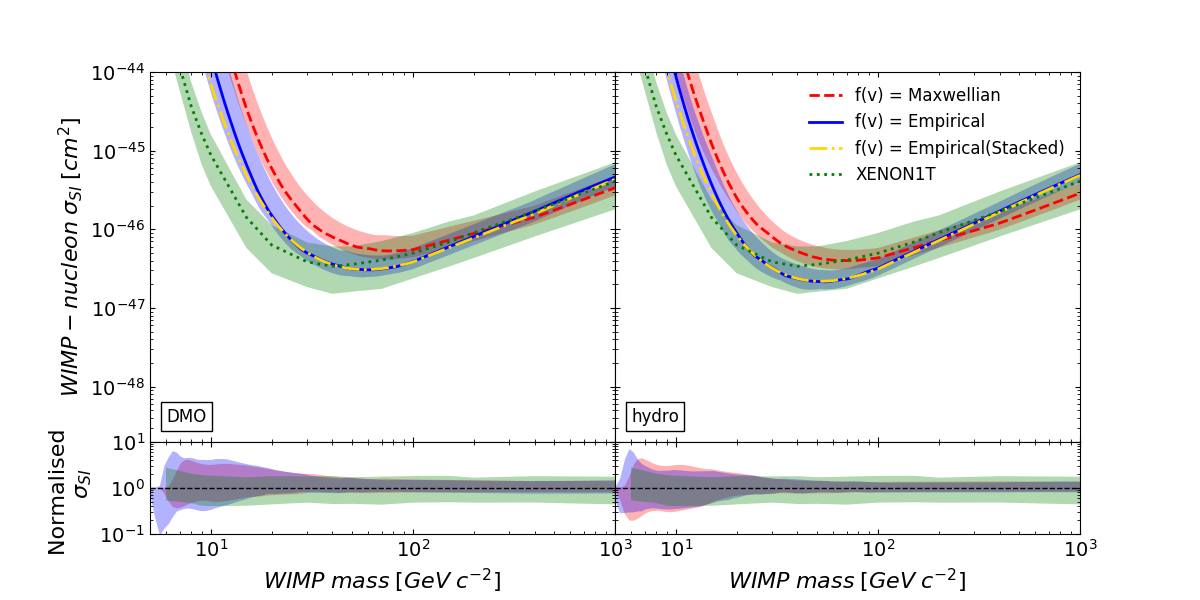}
  \caption{\textit{Left}: The 90\% confidence level limits using the \texttt{ARTEMIS} DMO halos using XENON1T parameters. The median upper-limits are shown as red dashed, and solid blue lines for a Maxwellian $f(v)$ and empirical $f(v)$, respectively.  The yellow dot dashed lines show the results from stacking all of the halos VDFs in our sample. The contours enclose 68\% of all the individual exclusion limits for all of the halos. The green dotted line shows the experimental limit from XENON1T along with the 1$\sigma$ error (green contour), which uses the Standard Halo Model. \textit{Right}: Same as left, but for the hydro halos. The lower panels show the uncertainties on the WIMP cross-sections normalised by their respective 90\% confidence level limits.}
  \label{fig:f(v)}
\end{figure}

\subsection{Empirical $f(v)$ model}
\label{sec:empirical}

We have shown that the Maxwellian function does not describe the DMO simulations well and, even though this issue is significantly reduced for the hydro simulations, it still does not represent a perfect description of the velocity distribution in those simulations either.  We therefore explore the impact of using the exact (empirical) form for $f(|v|)$ from the simulations to calculate the detection limits, via equation \ref{eq:BBRate}.  We calculate the detection limits for each halo, interpolating $f(|v|)$ where necessary between the $v_{\rm min}$ and $v_{\rm max}$ limits.

The results of using the empirical DM velocity distributions in both DMO and hydro simulations can be seen in the left and right panels of Fig.~\ref{fig:f(v)}, respectively.  The red dashed curve represents the median exclusion limits where $f(|v|)$ is assumed to be a Maxwellian distribution (as used in the SHM), and the blue solid line shows the median result when using the empirical $f(|v|)$ determined directly from the simulations. The yellow dot dashed line shows the limits placed on the WIMP cross-section using an empirical model derived from stacking all of the halos in our sample. Stacking the halos better populates the tails of the velocity distribution function, leading to a higher integrated rate and, therefore, lower WIMP masses to be excluded.  (This is similar to increasing the value of $v_{\rm esc}$.)

This comparison shows that the empirical method results in a significant reduction in the upper limits placed on the cross section for masses less than $\approx$ 100 GeV $c^{-2}$.  This difference is present for both the DMO and hydro cases but is larger for the DMO case due to its more significant deviation from the a Maxwellian distribution (particularly at high velocities). The halo-to-halo variation still persists in the empirical case and remains comparable to both the Maxwellian model and experimental results, reinforcing the importance $\rho_0$, $v_0$, $v_{\rm esc}$ when determining the exclusion limits.



\section{Summary and conclusions}
\label{sec:Conclusion}

Direct detection experiments require estimates of the local DM density and velocity distributions in order to place constraints on the DM particle mass and cross-section.  We have used the new \texttt{ARTEMIS} suite of high-resolution cosmological simulations of Milky Way-mass galaxies to determine the local densities and velocity distributions of DM in the presence or absence of baryons and across a variety of galaxy accretion histories. We have investigated the range in $\rho_0$, $v_0,$ $v_{\rm esc}$ in the simulations in order to inform the standard halo model implemented in direct detection pipelines. We have also investigated the degree to which the Maxwellian assumption for DM particle velocities is valid and have explored the impact of using a more realistic empirical function $f(v)$ from the simulations.  Using these results, we have estimated the uncertainties in the WIMP cross-section limits for the LZ and XENON1T direct detection experiments, under the assumption of a WIMP-nucleon spin-independent elastic scattering. 

Our main conclusions are as follows:

\begin{enumerate}[i.]
    \item The simulations predict local parameters ($\rho_0$, $v_0$, $v_{\rm esc}$) which are generally in good agreement with observations, within the observational uncertainties. The local DM density in the solar neighbourhood ranges between $0.15 - 0.48$ GeV cm$^{-3}$ in the hydro simulations and between $0.10 - 0.38$ GeV cm$^{-3}$ in the DMO simulations (Fig.~\ref{fig:LocDen}). The increased values of $\rho_0$ in the hydro simulations relative the DMO simulations are likely to be caused by the adiabatic contraction in the presence of baryons. Even taking this effect into account, our simulations disfavour the larger values of $>0.6$ GeV cm$^{-3}$ obtained in some recent measurements.
    
    \item The median local DM velocity distribution is relatively well (but not perfectly) described by a Maxwellian distribution for the hydro halos, but less so for the DMO halos (Fig.~\ref{fig:vmod}). Relatively large halo-to-halo variations are observed in both hydro and DMO simulations (Fig.~\ref{fig:App_VDF}). The addition of baryons and associated cooling and adiabatic contraction of the DM cause an increase in the peak velocities, typically by about 30 km s$^{-1}$.
    
    \item Substructure (subhalos, streams) can pass inside the solar neighbourhood (Fig.~\ref{fig:VDF_sub}), however this does not often occur in the cosmological context for relatively high-mass substructures that could significantly influence the DM detectability rates. Dark discs, however, are found in a relatively large proportion of the hydro halos. By a conservative estimate,  $\approx 15\%$ of our sample have dark discs, but the fraction can be as high as $\approx 36\%$ (Fig.~\ref{fig:Vphi_peak}) depending on how one defines a dark disc. The presence of dark discs increases the local DM density and can also lead to significant deviations from Maxwellian velocity distribution. 
    
    \item The enhanced DM densities and peak velocities in hydro simulations relative to DMO simulations lead to higher differential scattering rates in the former, and therefore to lower exclusion limits on the WIMP-nucleon cross-section (Fig.~\ref{fig:CS}).  In addition, the halo-to-halo scatter in the implied exclusion limits (due to scatter in the density and velocity distributions) is relatively large; typically a factor $\approx 1.5$ and increasing towards lower WIMP masses.  This is only slightly smaller than the experimental uncertainties on the published XENON1T and LZ (projected) limits.  In fact, at low WIMP masses ($< 20$ GeV $c^{-2}$) the simulation-based scatter typically exceeds the experimental uncertainty.
    We therefore conclude that the astrophysical systematic uncertainties should in general be included as part of the overall error budget.  This is important both for deriving a conservative estimate on the cross-section limits and for when comparisons are made to constraints on DM properties derived from indirect and collider searches. 
    
    \item An empirical form for $f(\vec{v})$ slightly lowers the exclusion limits, with the most significant difference seen when at WIMP mass $<$ 100 GeV $c^{-2}$ (Fig.~\ref{fig:f(v)}). This suggests that use of a Maxwellian form in the SHM generally puts a conservative limit on the exclusion limits compared with a more realistic model for the velocity distribution at masses of $< 100$ GeV $c^{-2}$ (at higher masses, the opposite is true).
    
\end{enumerate}

\acknowledgments

We thank the referee for their helpful and constructive comments. RPM acknowledges a LIV.DAT doctoral studentship supported by the STFC [ST/P006752/1]. The LIV.DAT Centre for Doctoral Training (CDT) is hosted by the University of Liverpool and Liverpool John Moores University / Astrophysics Research Institute. This project has received funding from the European Research Council (ERC) under the European Union's Horizon 2020 research and innovation programme (grant agreement No 769130). This work used the DiRAC@Durham facility managed by the Institute for Computational Cosmology on behalf of the STFC DiRAC HPC Facility. The equipment was funded by BEIS capital funding via STFC capital grants ST/P002293/1, ST/R002371/1 and ST/S002502/1, Durham University and STFC operations grant ST/R000832/1. DiRAC is part of the National e-Infrastructure.


\bibliographystyle{JCAP}
\bibliography{references}

\appendix
\section{Halo properties and velocity distribution functions}
\label{sec:appendix}

In this appendix, we provide the main properties of the \texttt{ARTEMIS} halos and their individual velocity distribution functions, for both the DMO and hydro cases.

Table \ref{tab:halo_properties} contains the tabulated halo properties of the simulated samples, including the spherical overdensity masses and radii, the maximum circular velocities, and the local DM densities (i.e., in a cylindrical shell at the solar radius, as described in the main text).

Table \ref{tab:chi2} contains the tabulated reduced chi-squared values of the Maxwellian velocity distributions for all halos in the DMO and hydro cases. The values clearly show that the velocity distributions of the hydro halos are better fitted by a Maxwellian distribution than the DMO halos.

\begin{figure}[hbt!]
  \centering    
  \includegraphics[width=\textwidth]{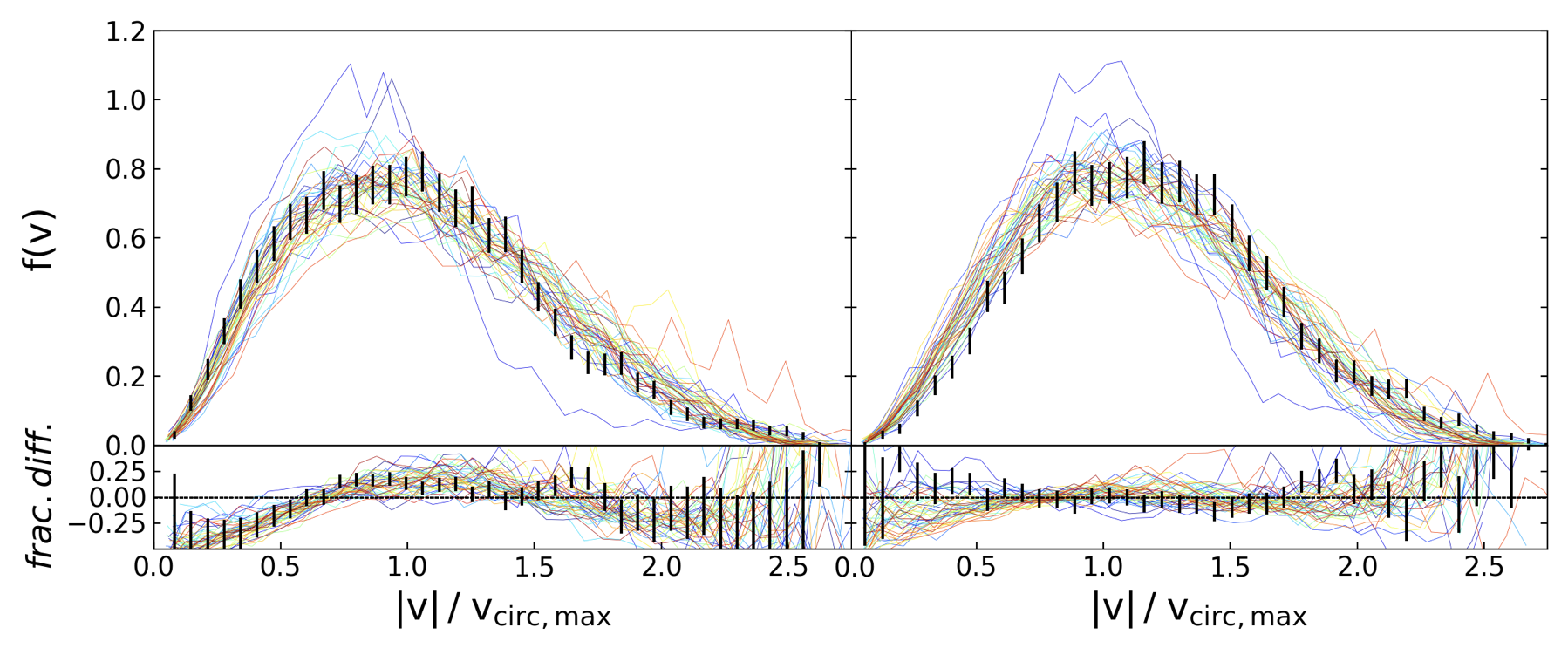}
  \caption{\textit{Left}: Local DM velocity modulus distributions in the rest frame of the galaxy, normalised by their respective maximum circular velocity, $\rm{v_{circ,max}}$. Coloured lines represent individual local velocity modulus distributions for all of the DMO halos. The black lines represent the 1$\sigma$ Poisson error for a single halo. The lower panel shows the fractional difference between the individual distributions and their best fit Maxwellian. \textit{Right}: Same as above but for the hydro halos.}
  \label{fig:App_VDF}
\end{figure}

Fig.~\ref{fig:App_VDF} shows the individual velocity distribution functions for the DMO (left panel) and hydro (right panel) cases.  The error bars show the Poisson errors for a representative halo.  The bottom panels show the difference of the best-fit Maxwellian with respect to the true velocity distribution function for that halo.  In general, the DMO halos are poorly described by a Maxwellian at low velocities.  Including hydrodynamics and galaxy formation leads to adiabatic contraction of the DM halo and a more isothermal distribution, with a Maxwellian yielding a better match.  Note, however, that for an individual halo deviations from the best-fit Maxwellian can occasionally exceed tens of percent in a given velocity bin (greatly exceeding the random error).

\begin{landscape}
\LTcapwidth=\linewidth
{\setlength{\tabcolsep}{2pt}
\begin{longtable} {crrccrrcccccrr}

\caption{The main properties of the Milky Way-analog halos in the \texttt{ARTEMIS} simulations for the DMO and hydro cases. The columns include: The ID name of the simulated galaxy, the virial mass ($\rm{M_{200}^{DMO/hydro}}$), the total stellar mass ($\rm{M_{*}^{hydro}}$), the virial radius ($\rm{R_{200}^{DMO/hydro}}$), the maximum circular velocity ($\rm{v_{circ,max}^{DMO/hydro}}$), the local dark matter density ($\rm{\rho_{0}^{DMO/hydro}}$), the local peak velocity ($\rm{v_{0}^{DMO/hydro}}$) and the hydro peak azimuthal velocities for DM and stars ($\rm{v_{\phi, peak, DM/stars}^{hydro}}$).}

\label{tab:halo_properties}
\\ \hline
Halo & $\rm{M_{200}^{DMO}}$ & $\rm{M_{200}^{hydro}}$ & $\rm{M_{*}^{hydro}}$ & $\rm{R_{200}^{DMO}}$ & $\rm{R_{200}^{hydro}}$ & $\rm{v_{circ,max}^{DMO}}$ & $\rm{v_{circ,max}^{hydro}}$ & $\rm{\rho_{0}^{DMO}}$ & $\rm{\rho_{0}^{hydro}}$ & $\rm{v_{0}^{DMO}}$ & $\rm{v_{0}^{hydro}}$ & $\rm{v_{\phi, peak, DM}^{hydro}}$ & $\rm{v_{\phi, peak, stars}^{hydro}}$\\

     & ($10^{11}M_{\odot}$) & ($10^{11}M_{\odot}$) & ($10^{10}M_{\odot}$) & (kpc) & (kpc)   & ($\rm{km \; s^{-1}}$)       & ($\rm{km \; s^{-1}}$)         & ($\rm{GeV \; cm^{-3}}$)      & ($\rm{GeV \; cm^{-3}}$)    & ($\rm{km \; s^{-1}}$)       & ($\rm{km \; s^{-1}}$)    & ($\rm{km \; s^{-1}}$)       & ($\rm{km \; s^{-1}}$)\\

\hline
\endfirsthead

 & & & & & & & & & & & & & continued...\\
\endfoot
\endlastfoot

\\ \hline
Halo & $\rm{M_{200}^{DMO}}$ & $\rm{M_{200}^{hydro}}$ & $\rm{M_{*}^{hydro}}$ & $\rm{R_{200}^{DMO}}$ & $\rm{R_{200}^{hydro}}$ & $\rm{v_{circ,max}^{DMO}}$ & $\rm{v_{circ,max}^{hydro}}$ & $\rm{\rho_{0}^{DMO}}$ & $\rm{\rho_{0}^{hydro}}$ & $\rm{v_{0}^{DMO}}$ & $\rm{v_{0}^{hydro}}$ & $\rm{v_{\phi, peak, DM}^{hydro}}$ & $\rm{v_{\phi, peak, stars}^{hydro}}$\\

     & ($10^{11}M_{\odot}$) & ($10^{11}M_{\odot}$) & ($10^{10}M_{\odot}$) & (kpc) & (kpc)   & ($\rm{km \; s^{-1}}$)       & ($\rm{km \; s^{-1}}$)         & ($\rm{GeV \; cm^{-3}}$)      & ($\rm{GeV \; cm^{-3}}$)    & ($\rm{km \; s^{-1}}$)       & ($\rm{km \; s^{-1}}$)    & ($\rm{km \; s^{-1}}$)       & ($\rm{km \; s^{-1}}$)\\

\hline
\endhead
G1  & 12.50 & 11.90 & 3.64 & 222.24 & 218.59 & 178.84 & 199.44 & 0.28 & 0.45 & 174.34 & 204.88 & 95.00  & 193.00\\
G2  & 17.17 & 16.53 & 3.87 & 247.02 & 243.89 & 185.12 & 189.85 & 0.23 & 0.29 & 161.07 & 196.50 & 69.00  & 151.00\\
G3  & 19.17 & 17.01 & 3.92 & 256.26 & 246.25 & 194.29 & 204.20 & 0.21 & 0.24 & 180.02 & 211.46 & 33.00  & 195.00\\
G4  & 18.01 & 14.32 & 3.32 & 250.95 & 232.49 & 186.83 & 181.22 & 0.24 & 0.30 & 173.63 & 198.04 & 73.00  & 145.00\\
G5  & 18.87 & 16.42 & 3.25 & 254.89 & 243.36 & 171.98 & 186.22 & 0.10 & 0.15 & 133.23 & 165.21 & 11.00  & 147.00\\
G6  & 18.30 & 16.44 & 5.45 & 252.31 & 243.47 & 199.41 & 230.25 & 0.20 & 0.26 & 207.56 & 218.86 & 53.00  & 163.00\\
G7  & 10.62 & 9.97  & 2.26 & 210.45 & 206.05 & 156.00 & 177.20 & 0.17 & 0.24 & 157.81 & 170.98 & 3.00   & 169.00\\
G8  & 18.77 & 16.31 & 2.19 & 254.44 & 242.80 & 185.69 & 185.43 & 0.21 & 0.23 & 179.20 & 170.20 & 3.00   & 213.00\\
G9  & 11.48 & 11.09 & 3.73 & 216.02 & 213.53 & 165.72 & 187.36 & 0.26 & 0.29 & 149.82 & 189.77 & 107.00 & 167.00\\
G10 & 13.64 & 11.53 & 2.42 & 228.79 & 216.28 & 195.80 & 188.50 & 0.29 & 0.33 & 192.61 & 210.10 & 5.00   & 177.00\\
G11 & 13.86 & 11.69 & 4.13 & 229.99 & 217.28 & 176.79 & 179.60 & 0.21 & 0.31 & 161.80 & 181.72 & 77.00  & 143.00\\
G12 & 14.28 & 13.23 & 3.94 & 232.29 & 226.48 & 173.65 & 175.19 & 0.22 & 0.22 & 156.10 & 178.91 & -19.00 & 161.00\\
G13 & 11.68 & 11.69 & 2.17 & 217.21 & 217.28 & 142.84 & 153.44 & 0.10 & 0.21 & 145.29 & 152.09 & 17.00  & 133.00\\
G14 & 14.98 & 12.15 & 3.53 & 236.02 & 220.14 & 206.92 & 225.66 & 0.30 & 0.38 & 220.32 & 231.47 & 27.00  & 211.00\\
G15 & 11.81 & 11.22 & 3.57 & 218.06 & 214.32 & 155.14 & 170.19 & 0.17 & 0.20 & 134.23 & 174.81 & 21.00  & 155.00\\
G16 & 12.22 & 12.69 & 2.94 & 220.56 & 223.31 & 175.39 & 175.54 & 0.30 & 0.33 & 160.41 & 169.31 & 61.00  & 171.00\\
G17 & 12.55 & 11.69 & 3.74 & 222.48 & 217.28 & 188.81 & 198.05 & 0.32 & 0.43 & 179.49 & 203.64 & 69.00  & 191.00\\
G18 & 13.12 & 9.68  & 2.78 & 225.82 & 204.03 & 175.11 & 183.74 & 0.31 & 0.40 & 170.40 & 198.38 & -35.00 & 173.00\\
G19 & 10.77 & 9.62  & 2.57 & 211.43 & 203.67 & 176.59 & 176.88 & 0.26 & 0.35 & 176.16 & 191.02 & 17.00  & 175.00\\
G20 & 11.06 & 10.58 & 3.36 & 213.30 & 210.16 & 171.74 & 184.52 & 0.28 & 0.38 & 158.01 & 189.52 & 61.00  & 143.00\\
G21 & 12.17 & 10.11 & 1.75 & 220.23 & 207.06 & 169.32 & 160.90 & 0.27 & 0.27 & 165.48 & 175.84 & -7.00  & 107.00\\
G22 & 11.81 & 10.08 & 2.87 & 218.03 & 206.84 & 177.64 & 178.89 & 0.19 & 0.29 & 180.56 & 204.54 & 65.00  & 171.00\\
G23 & 11.11 & 9.95  & 2.87 & 213.62 & 205.93 & 167.66 & 196.69 & 0.26 & 0.34 & 157.81 & 195.58 & 1.00   & 187.00\\
G24 & 11.15 & 10.29 & 3.63 & 213.90 & 208.28 & 165.04 & 185.29 & 0.17 & 0.31 & 167.51 & 199.75 & -33.00 & 187.00\\
G25 & 9.25  & 9.12  & 2.58 & 200.97 & 200.02 & 166.99 & 171.76 & 0.34 & 0.34 & 148.95 & 170.49 & 57.00  & 169.00\\
G26 & 10.32 & 8.96  & 3.53 & 208.43 & 198.86 & 173.33 & 195.18 & 0.23 & 0.35 & 175.87 & 203.86 & 33.00  & 183.00\\
G27 & 8.48  & 7.96  & 2.57 & 195.27 & 191.18 & 149.06 & 159.55 & 0.18 & 0.30 & 150.74 & 169.26 & 9.00   & 153.00\\
G28 & 8.09  & 7.67  & 2.39 & 192.21 & 188.83 & 137.26 & 165.97 & 0.15 & 0.24 & 140.91 & 167.33 & -3.00  & 157.00\\
G29 & 9.99  & 8.82  & 3.11 & 206.18 & 197.87 & 168.53 & 210.45 & 0.29 & 0.41 & 170.61 & 201.80 & 1.00   & 201.00\\
G30 & 9.29  & 8.08  & 2.69 & 201.26 & 192.16 & 168.58 & 171.58 & 0.27 & 0.33 & 167.90 & 193.31 & 37.00  & 163.00\\
G31 & 8.73  & 8.32  & 2.09 & 197.14 & 193.99 & 153.77 & 160.32 & 0.31 & 0.32 & 145.25 & 158.02 & 55.00  & 141.00\\
G32 & 7.93  & 7.88  & 2.51 & 190.94 & 190.51 & 150.29 & 155.27 & 0.18 & 0.23 & 141.24 & 162.24 & 7.00   & 151.00\\
G33 & 9.15  & 7.80  & 2.64 & 200.27 & 189.93 & 158.48 & 163.05 & 0.29 & 0.32 & 141.62 & 165.70 & 19.00  & 139.00\\
G34 & 8.76  & 7.89  & 2.85 & 197.34 & 190.65 & 170.13 & 183.40 & 0.32 & 0.45 & 162.64 & 181.81 & -11.00 & 181.00\\
G35 & 7.99  & 6.82  & 1.91 & 191.42 & 181.57 & 160.06 & 164.13 & 0.27 & 0.35 & 163.69 & 178.41 & 15.00  & 155.00\\
G36 & 38.57 & 36.36 & 4.49 & 323.49 & 317.19 & 219.36 & 214.47 & 0.13 & 0.18 & 249.83 & 237.47 & 19.00  & 181.00\\
G37 & 8.44  & 6.66  & 1.76 & 194.95 & 180.11 & 161.56 & 162.61 & 0.32 & 0.36 & 155.79 & 167.95 & -5.00  & 143.00\\
G38 & 8.17  & 7.14  & 2.97 & 192.86 & 184.35 & 157.22 & 175.70 & 0.25 & 0.48 & 150.58 & 173.85 & 61.00  & 171.00\\
G39 & 8.50  & 7.48  & 1.88 & 195.37 & 187.24 & 167.63 & 165.57 & 0.38 & 0.39 & 168.29 & 175.31 & 21.00  & 149.00\\
G40 & 8.03  & 7.57  & 2.02 & 191.75 & 187.99 & 132.99 & 154.89 & 0.17 & 0.26 & 116.82 & 160.30 & 13.00  & 147.00\\
G41 & 8.12  & 6.89  & 1.94 & 192.40 & 182.18 & 166.01 & 161.58 & 0.35 & 0.41 & 158.66 & 158.51 & 101.00 & 139.00\\
G42 & 8.22  & 7.18  & 2.31 & 193.26 & 184.68 & 164.77 & 174.25 & 0.34 & 0.39 & 157.90 & 186.80 & 51.00  & 171.00\\

\hline

\end{longtable}
}
\end{landscape}

\section{Experimental corrections to the calculated exclusion limits}
\label{sec:appendixLimits}

\begin{figure}[hbt!]
  \centering    
  \includegraphics[width=\textwidth]{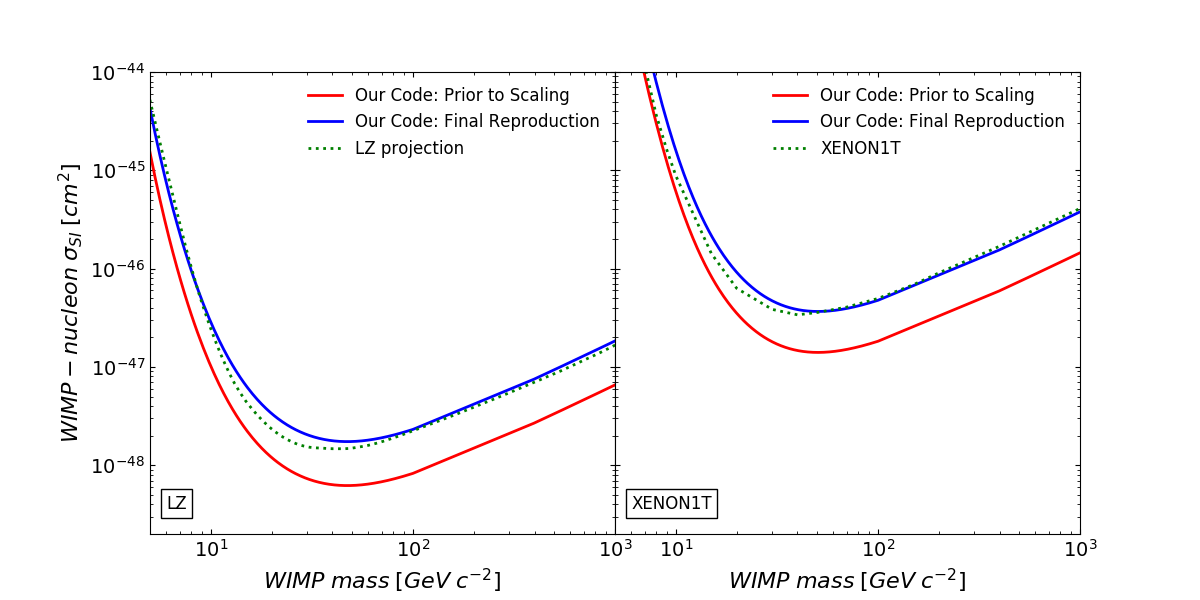}
  \caption{\textit{Left}: The 90\% confidence level limits for LZ parameters \cite{LZ_2018}.  The median upper-limits are shown as red solid and blue solid lines for before and post scaling to 100 GeV $c^{-2}$ respectively. The green dotted line shows the published experimental projections for LZ. All three lines assuming the SHM. \textit{Right}: Same as left, but now using the XENON1T experimental parameters to calculate the 90\% confidence level limits.}
  \label{fig:Detector_lims}
\end{figure}

Data analysis and limit setting procedures used by the direct detection experiments involve many steps and detector knowledge which can not be fully replicated outside of the experiments. To make a comparison of the calculated and experimental limits meaningful, especially at low masses where the effect of the velocity model is strongest, we included the recoil energy-dependent efficiency published by the experiments into the calculation. This efficiency affects the shape of the exclusion limit curve through the dependence of the recoil energy on the mass of dark matter particle. Still, as can be seen in Fig.~\ref{fig:Detector_lims}, the calculated and experimental limits do not agree. Additional signal region selection efficiencies, as well as event reconstruction efficiencies, should be included. These efficiencies are determined by experiments using full detector simulations. Their dependence on the DM particle mass is not strong, and they could be taken into account as simple scale factors at high masses (we take 100 GeV $c^{-2}$ mass point). The resulting limits agree with the experimental ones much better. The remaining discrepancies could be due to non-perfect matching of efficiency curves extracted from the publications and possible mass dependence of signal region selection and event reconstruction efficiencies. Given we are interested in relative effects due to velocity models, this agreement is good enough for our purposes. One disadvantage of scaling is that it is no longer possible to see the impact of the local density differences between the calculated and experimental limits.

Fig.~\ref{fig:Detector_lims} shows the 90$\%$ confidence limit on the SI WIMP-nucleon cross-section as taken from the relevant published article and the produced value of the code before and post scaling for both LZ ({\it left}) and XENON1T ({\it right}). When reproducing the limit, the SHM has been used with velocities as stated by the relevant detectors publication.

\begin{table}[!htbp]
\centering

 \caption{The goodness of fit values of the Maxwellian velocity distributions for all halos in the DMO and hydro cases ($\rm{\chi_{v}^{2 \;  {\rm DMO}}}$ and $\rm{\chi_{v}^{2 \; {\rm hydro}}}$, respectively).}
 \begin{tabular}{lrc | lrc}
 \\ \hline
 Halo & $\chi_v^{2 \; {\rm DMO}}$ & $\chi_v^{2 \; {\rm hydro}}$ & Halo & $\chi_v^{2 \; {\rm DMO}}$ & $\chi_v^{2 \; {\rm hydro}}$ \\ 
 \hline
 G1  & 7.07  & 1.07 & G22  & 10.35 & 2.76 \\ 
 G2  & 4.76  & 1.63 & G23  & 4.37  & 1.32 \\
 G3  & 5.44  & 1.95 & G24  & 5.76  & 0.93 \\
 G4  & 4.17  & 0.75 & G25  & 3.82  & 0.66 \\
 G5  & 36.51 & 5.44 & G26  & 4.62  & 1.59 \\
 G6  & 7.30  & 0.92 & G27  & 8.29  & 3.45 \\
 G7  & 1.83  & 3.07 & G28  & 11.76 & 2.18 \\
 G8  & 4,59  & 2.72 & G29  & 3.55  & 2.14 \\
 G9  & 3.47  & 3.40 & G30  & 9.63  & 2.05 \\
 G10 & 4.74  & 2.18 & G31  & 3.67  & 0.62 \\
 G11 & 3.62  & 0.71 & G32  & 2.51  & 1.81 \\
 G12 & 4.64  & 1.43 & G33  & 3.59  & 0.88 \\
 G13 & 1.71  & 1.50 & G34  & 5.09  & 3.70 \\
 G14 & 1.89  & 1.78 & G35  & 4.17  & 2.79 \\
 G15 & 10.37 & 1.75 & G36  & 5.77  & 4.56 \\
 G16 & 2.61  & 0.46 & G37  & 2.79  & 1.57 \\
 G17 & 3.33  & 0.71 & G38  & 1.22  & 0.97 \\
 G18 & 3.50  & 1.23 & G39  & 4.10  & 2.17 \\
 G19 & 2.13  & 1.94 & G40  & 10.75 & 1.72 \\
 G20 & 5.88  & 1.08 & G41  & 4.12  & 0.84 \\
 G21 & 3.98  & 0.70 & G42  & 4.21  & 1.61 \\
 \hline
 \label{tab:chi2}
\end{tabular}
\end{table}

\end{document}